# Experimental and theoretical study of stable and metastable phases in sputtered CuInS$_2$

*J. K. Larsen[1,§,*], K. V. Sopiha[1,§,*], C. Persson[2,3], C. Platzer-Björkman[1], M. Edoff[1]*

[1] Division of Solar Cell Technology, Department of Materials Science and Engineering, Uppsala University, Box 534, SE-75237 Uppsala, Sweden

[2] Centre for Materials Science and Nanotechnology/Department of Physics, University of Oslo, Blindern, Box 1048, NO-0316 Oslo, Norway

[3] Department of Materials Science and Engineering, Royal Institute of Technology, SE-10044 Stockholm, Sweden

* Corresponding authors: J. K. Larsen (jes.larsen@angstrom.uu.se), K. V. Sopiha (kostiantyn.sopiha@angstrom.uu.se)

[§] These authors contributed equally to the work

# Abstract

The chalcopyrite Cu(In,Ga)S$_2$ has gained renewed interest in recent years due to the potential application in tandem solar cells. In this contribution, a combined theoretical and experimental approach is applied to investigate stable and metastable phases forming in CuInS$_2$ (CIS) thin films. *Ab initio* calculations are performed to obtain formation energies, X-ray diffraction patterns, and Raman spectra of CIS polytypes and related compounds. Multiple CIS structures with zinc-blende and wurtzite-derived lattices are identified and their XRD/Raman patterns are shown to contain overlapping features, which could lead to misidentification. Thin films with compositions from Cu-rich to Cu-poor are synthesized with a two-step approach based on sputtering from binary targets followed by high-temperature sulfurization. It is discovered that several CIS polymorphs are formed when growing the material with this approach. In the Cu-poor material, wurtzite CIS is observed for the first time in sputtered thin films along with chalcopyrite CIS and CuAu-ordered CIS. Once the wurtzite CIS phase has formed, it is difficult to convert into the stable chalcopyrite polymorph. CuIn$_5$S$_8$ and NaInS$_2$ accommodating In-excess are found alongside the CIS polymorphs. It is argued that the metastable polymorphs are stabilized by off-stoichiometry of the precursors, hence tight composition control is required.

## 1. Introduction

The current world record for power conversion efficiency of Cu(In,Ga)(Se,S)$_2$-based solar cells was obtained by utilizing a "sulfurization-after-selenization" process [1]. This is done to incorporate sulfur in the surface of the absorber and widen the band gap near the junction with the buffer layer. The resulting record efficiency of 23.35% is not far behind the world record for crystalline Si solar cells, which is currently 26.6% - only a few percentage points below the theoretical limit of 29.1% for Si [2]. An established way of increasing solar cell efficiency beyond the current level is by making tandem or multi-junction devices. For a four-terminal tandem with a silicon bottom cell, it is theoretically possible to achieve efficiencies around 44% using a top cell with a band gap in the range 1.6−2.0 eV [3]. One candidate material for top cells in combination with Cu(In,Ga)Se$_2$ or Si is the sulfide Cu(In,Ga)S$_2$. Its band gap can be tuned from 1.5 eV for CuInS$_2$ to 2.4 eV for CuGaS$_2$, which covers the desired range for top cell applications. Historically, sulfide Cu(In,Ga)S$_2$ materials have received less attention than the more established selenide Cu(In,Ga)Se$_2$ because they are less suited for the traditional single-junction photovoltaics. The interest in the sulfides is, however, increasing with the enhanced focus on tandem solar cells. Early research has primarily focused on Cu-rich material [4], because conductivity was always too low for Cu-poor films [5], [6], but after initial advances the record efficiency for Cu-rich Cu(In,Ga)S$_2$ had saturated at around 13% despite extensive academic and industrial efforts [7]. Recently, a breakthrough was reported for Cu(In,Ga)S$_2$ solar cells reaching a conversion efficiency of 15.5% using Cu-poor material [8], which raises questions about the best processing route. This was further supported in a recent publication demonstrating 15.2% efficiency for a Cu(In,Ga)S$_2$-based device by reducing the [Cu]/([Ga]+[In]) ratio and utilizing a Zn(O,S) buffer layer to supress interface recombination [9]. Motivated by the improved efficiency, we are set to revise the details of growth chemistry in Cu-poor CuInS$_2$ (CIS). Besides solar cells, the obtained results might also be of interest for thermoelectric applications [10], [11] and beyond.

The first complete pseudo-binary phase diagram of Cu$_2$S-In$_2$S$_3$ system was published in 1980 by Binsma et al. [12] (see Figure 1). It shows that CuInS$_2$ exists in the chalcopyrite phase with a tetragonal structure (*CH-CIS*; space group *I-42d*) at room temperature and transforms into disordered zinc-blende-like sphalerite phase (*SPH-CIS;* space group *F-43m*) at 980 ºC, followed by conversion into disordered wurtzite-like phase (*WZ-CIS;* space group *P6$_3$mc*) at 1045 °C. Both disordered phases are characterised by random occupation of cationic sites while the anionic sublattice is filled by S atoms. Unlike many I-III-VI$_2$ isomorphs, chalcopyrite CuInS$_2$ has surprisingly narrow homogeneity region – only a few percent toward the In$_2$S$_3$-rich side. Thiospinel CuIn$_5$S$_8$ and Cu$_2$S form at the In-rich and Cu-rich compositions, respectively. Interestingly, the phase diagram also shows that slightly off-stoichiometric CIS remain disordered at up to 100 °C lower temperature compared to stoichiometric CuInS$_2$, suggesting that *SPH-CIS* and *WZ-CIS* phases can be stabilized by either excess or deficiency of Cu$_2$S. Asimilar stabilization effect was noted in selenide Cu-In-Se system, where the *SPH* phase could be observed down to room temperature when the composition fell between 15.5 and 11.0 at % Cu on the CuInSe$_2$-In$_{0.42}$Se$_{0.58}$ tie-line (considerably Cu-poor and Se-enriched as compared to the chalcopyrite phase) [13]. All these findings highlight the intricate interplay between stoichiometry and stability of CuInS$_2$ polymorphs, which needs to be duly accounted for during the deposition of solar absorbers.

The renewed interest in Cu(In,Ga)S$_2$ absorbers stimulated a detailed study of the crystalline phases observed in the Cu$_2$S-In$_2$S$_3$-Ga$_2$S$_3$ pseudo-ternary system in 2018 [14]. In that study, Thomere et al. point out that sulfide Cu(In,Ga)S$_2$ is much more structurally complex than its selenide counterpart. It is particularly important that the sulfide chalcopyrite structure is less adaptable to Cu-deficiency, reflecting the narrow homogeneity range reported by Binsma et al. [12]. They concluded that any attempt to produce Cu-poor Cu(In,Ga)S$_2$ in sealed silica tubes at 800 ºC resulted in formation of mixed chalcopyrite and Cu-poor phases such as the thiospinel CuIn$_5$S$_8$. C. Stephan investigated the extension of the single-phase region of CIS and observed CuIn$_5$S$_8$ precipitates together with *CH-CIS* already for [Cu]/[In] < 0.95, while *CH-CIS* and Cu$_x$S were found to co-exist for [Cu]/[In] > 1 [15], also in accordance with the phase diagram [12]. It should be noted that these studies were performed on powder samples produced under nearly equilibrium conditions (prolonged annealing at 800 °C or above in sealed quartz tubes followed by slow cooling [14], [15]), with the composition being intentionally set on the Cu$_2$S-In$_2$S$_3$ tie-line by controlling the elemental precursor ratios. In contrast, solar absorber films are typically deposited under the conditions far from equilibrium, with S deficiency or excess being possible during the growth, thereby opening the possibility to form metastable phases.

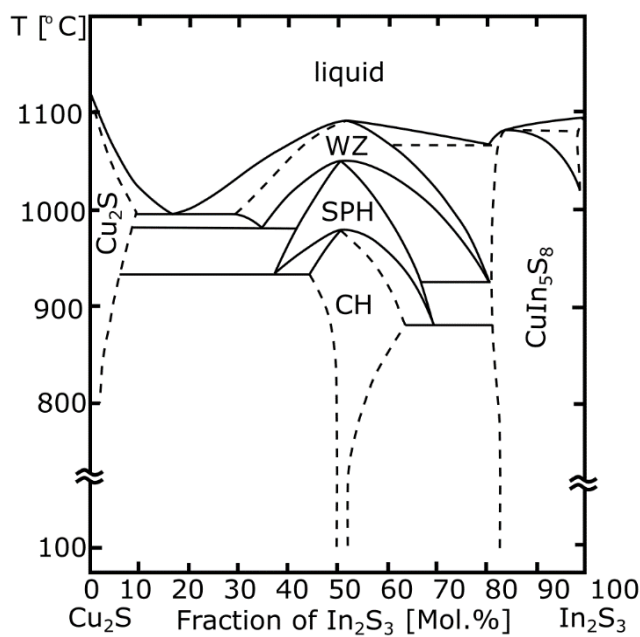

*Figure 1. Pseudo-binary phase diagram of $Cu_2S$-$In_2S_3$ system adapted from Binsma et al. [12].*

Apart from disordered *SPH-CIS* and *WZ-CIS*, and the ordered ground state *CH-CIS* phase, three other, potentially metastable ordered structures have been described. These are a CuAu-type structure based on a zinc-blende lattice (*ZB-CA;* space group *P-4m2*), as well as two structures based on the wurtzite lattice with chalcopyrite-type (*WZ-CH;* space group *Pna2$_1$*) and CuAu-type ordering (*WZ-CA;* space group *Pmc2$_1$*) [10], [16]. Figure 2 illustrates how the disordered phases transform into either the CH-like or CA-like structure upon ordering. All four ordered structures obey Grimm-Sommerfeld rule [17], often referred to as octet rule, which is a known prerequisite of low energy [18], [19]. In principle, infinite number of polytypes can be derived without violating the octet rule by, for example, incorporating antisite domain boundaries into *CH-CIS* [20], [21] or choosing a different mixing pattern for the WZ lattice [11]. Some of those structures were reproduced and analysed computationally in this work.

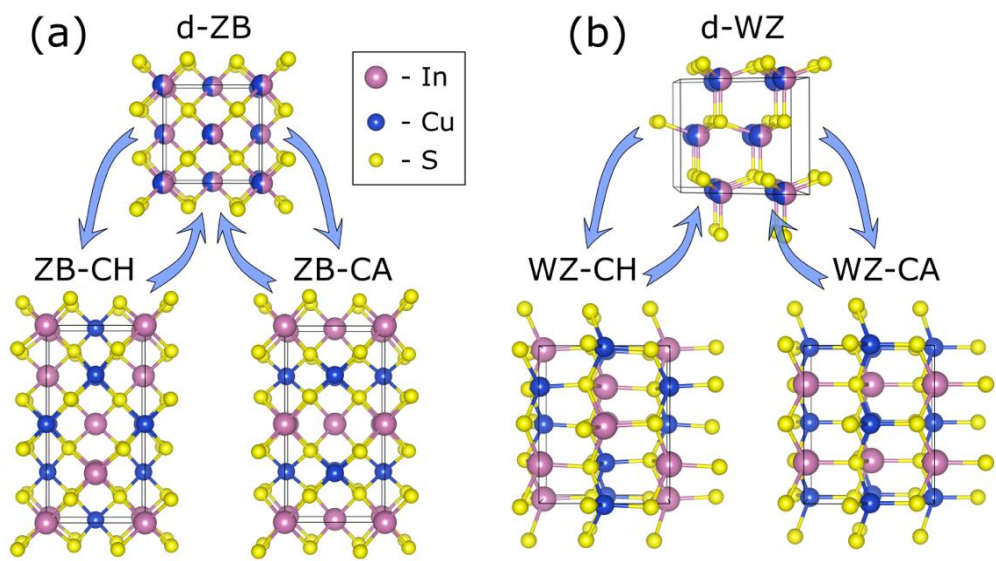

*Figure 2. Structures of the (a) ZB- and (b) WZ-based CuInS$_2$ phases discussed in this work. The structures labelled "d-" are the disordered phases. The CH-like and CA-like structures are obtained through cationic ordering, as indicated with the arrows.*

It is important to differentiate between "phases" and "structures" as these terms do not necessarily refer to the same thing – disordered phases can often be represented as statistical ensembles of inequivalent structural polytypes, or as structures with fractional site occupancies, whereas ordered structures may or may not be recognized in experiments. The crystal structures of the *CH-CIS*, *CA-CIS*, and *SPH-CIS* phases are known to be ZB-CH, ZB-CA, and d-ZB, respectively. Thus, these notations are equivalent and will be used when referring to either structures or phases in this paper. The structure of the *WZ-CIS* phase, in contrast, has not yet been fully resolved, necessitating to make distinctions between WZ-CH, WZ-CA, and d-WZ.

With such a diversity of structures, one should not be surprised of the lacking consensus in literature about the types and properties of metastable CIS phases. The main practical problem is that, owing to the structural similarity, all ZB-derived structures produce overlapping diffraction peaks, making it difficult to unambiguously identify them by XRD. *CH-CIS* has only a couple of unique reflections, most notably the (011) peak at $2\theta \approx 17.9°$ when using Cu K$\alpha$ radiation [22], whereas *CA-CIS* has unique (001) reflection at $2\theta \approx 16.1°$ and (100) reflection at $2\theta \approx 22.6°$, both of which are not seen in the other CIS polymorphs [23]–[25]. For reference, the respective diffraction patterns are given in Figure 3. In principle, these features should be enough to detect *CH-CIS* or *CA-CIS* in samples containing *SPH-CIS* but insufficient to rule out the presence of *SPH-CIS* in a mixture with either *CH-CIS* or *CA-CIS*. Moreover, in practice, the unique XRD reflections for ZB-derived phases are rarely strong enough to detect in thin films or small nanocrystals, rendering them unreliable for phase identification. Still, XRD has been adopted as the primary tool to conclude that polycrystalline powder produced via high-pressure solid-state reaction [23] or epitaxially grown films via molecular beam epitaxy on silicon (001) substrate [24] can be composed predominantly of the *CA-CIS* polymorph. It should be noted though that there is also a number of reports suggesting that epitaxial [25]–[28] and co-evaporated [29], [30] films can also consist of intermixed *CH-CIS* and *CA-CIS*. Another notable observation in literature is that the fraction of *CA-CIS* phase increases when lowering the [Cu]/[In] ratio in co-evaporated films [29], [30], suggesting that it can tolerate higher off-stoichiometry than *CH-CIS*.

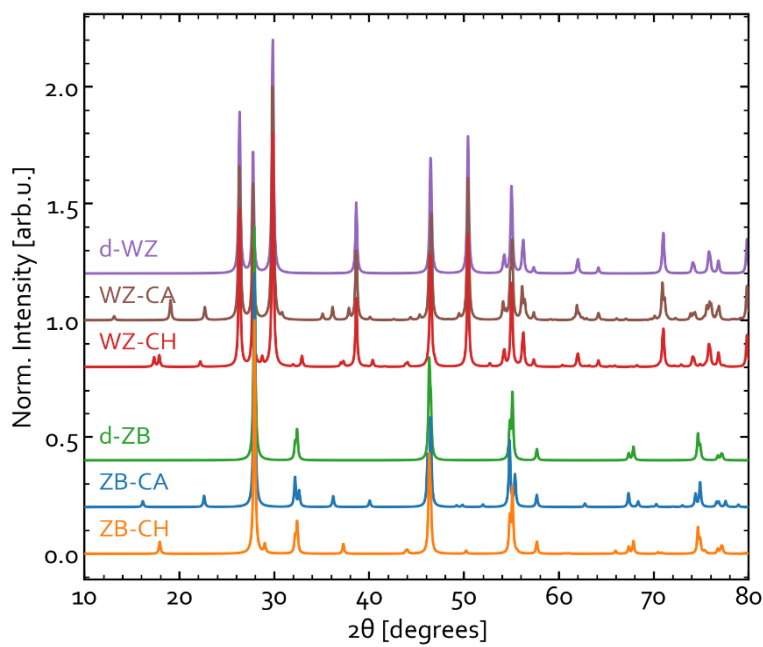

*Figure 3. Simulated XRD patterns for different CIS structures (under Cu Kα radiation). The patterns were simulated using XRD diffraction module in pymatgen, which is based on the formalism laid out elsewhere [31]. To add spectral broadening, Lorentzian smearing with FWHM of 0.2° was applied to all patters presented.*

Different *WZ*-derived structures produce similar diffraction patterns but all distinct from those of ZB-derived CIS (see Figure 3). This was first noted experimentally in 1998 by Schimmel et al. [32], who detected *WZ-CIS* in electrodeposited CIS nanoparticles by XRD. However, it was not until 2008 when Pan et al. [33] discovered a simple way to toggle between *ZB-* and *WZ*-derived phases by changing the capping agent in hot-injection synthesis that *WZ-CIS* started to be reliably produced, primarily via low-temperature (below 300 ºC) wet-chemical routes [11], [22], [34]–[38]. Another parameter that was claimed to trigger preferential growth of *WZ-CIS* is the solutions pH [37]. Such phase tuning indicates that *WZ-CIS* is stabilized by lowering surface energy in the solution during growth. This effect must be particularly important for nanocrystals due to their high specific surface area [10]. At the same time, the smaller contribution of surface energy in thin films justifies why no earlier study has detected *WZ-CIS* precipitation in sputtered CIS films.

A handful of experimental evidence exists that the cationic disorder vanishes at room temperature. First, Binsma et al. [12] noted that freezing *SPH-CIS* and *WZ-CIS* by quenching in liquid $N_2$ is impossible. Instead, using Monte-Carlo simulations based on the energies of $CuInSe_2$ structures computed using the density functional theory (DFT), Wei et al. [18] concluded that disordered I-III-$VI_2$ at room temperature attain mixed *CH-* and *CA*-type features if the material is grown below the critical order-disorder transition temperature. Subsequently, the presence of *CA-CIS* domains in epitaxially grown CIS film was detected with high-resolution transmission electron microscopy [26], [27], [39] shortly thereafter. Likewise, Shen et al. [11] found that *WZ-CIS* nanocrystals grown by a solution-phase colloidal synthesis are composed of interlaced domains of ordered wurtzite structures. This conclusion was later confirmed elsewhere [40].

Raman spectroscopy is more effective than XRD when it comes to detection of *CA-CIS*. It was proven that *CA-CIS* has a $A_1$-S-S vibration mode at about 5% higher wavenumbers than *CH-CIS*, appearing at about 305 cm$^{-1}$ for *CA-CIS* as compared to 292 cm$^{-1}$ for *CH-CIS* [39]. At the same time, untangling Raman signatures for the *WZ-CIS* phases is ambiguous – a few benchmark studies exist [41], [42] but they all show several peaks that may overlap with the $A_1$-S-S modes of ZB-derived CIS and/or other secondary phases, such as $CuIn_5S_8$. Dzhagan et al. suggested that the key identifier of *WZ-CIS* is a peak at 340-350 cm$^{-1}$, which is found to be sensitive to the choice of excitation wavelength [41]. An alternative explanation is that this signal stems from $B_2^{(3)}(L)/E^{(6)}(L)$ mode of *CH-CIS* experiencing quasi-resonance enhancement under excitation by infrared laser ($\lambda_{exc}$=785 nm) [43], [44]. As a further complication for phase identification is that, to our best knowledge, neither computed nor measured reference Raman spectra exists for disordered *SPH-CIS* and *WZ-CIS* in literature.

In this work, a joint experimental and theoretical approach is applied in order to enhance understanding of metastable CIS polymorphs. *Ab initio* DFT calculations are performed to obtain formation energies, XRD patterns, and theoretical Raman spectra of CIS polymorphs and related In-rich phases. The results are compared to experimental work in which a two-step CIS fabrication process is adopted that relies on 1) compound sputtering of a Cu-In-S precursor and 2) subsequent annealing in a sulfur-containing atmosphere. A similar approach to CIS preparation has been explored in the past (see for example [5], [45]), while an alternative one-step sputtering route is likewise commonly employed (*e.g.* [46], [47]). The produced films are thoroughly characterised and found to contain varying amounts of CIS phases depending on the precursor composition and processing conditions. Based on this data, crystal structures and growth tendencies for metastable CIS phases are assessed and discussed.

# 2. Methods

## 2.1 Sample preparation

### 2.1.1 Sputtering and annealing of Cu-In-S

Prior to CIS deposition, soda-lime glass (SLG) substrates were coated with about 350 nm Mo by DC-magnetron sputtering. A series of Cu-In-S precursors were deposited onto Mo by co-sputtering from $In_2S_3$ and CuS binary targets without intentional substrate heating. Samples with different [Cu]/[In] ratios were prepared by variation of the relative power of the targets and the resulting compositions of the precursors were measured by X-ray fluorescence (XRF). For some samples, 20 nm NaF was deposited onto the precursors by evaporation before annealing. The precursors were subsequently annealed at various temperatures in a tube furnace described in more details in reference [48]. For the annealing the precursors were loaded in a pyrolytic carbon coated graphite box containing 180 mg of elemental sulfur. The graphite box was introduced into the preheated furnace with an argon background pressure of 350 Torr. After samples were transferred into the hot zone, the temperature of the box increased to the target temperature in about 90 s. The samples were then allowed to dwell for 30 min before the graphite box was removed from the hot zone, resulting in a rapid cool down in the Ar atmosphere. The reported sample temperature was measured at the base of the graphite box. No sulfur remained in the box after annealing.

#### 2.1.2 Reference material preparation

**A NaInS$_2$ thin-film reference sample** was produced in order to investigate the characteristic Raman peaks of the material. The sample was grown on soda-lime glass by first depositing 50 nm of NaF using evaporation, followed by thermal evaporation of 30 nm of metallic indium. The resulting precursor is estimated to have [Na]/[In] close to 1.5, ensuring substantial Na excess. In order to sulfurize the material, the precursor was loaded in a graphite carrier with 20 g of elemental sulfur and introduced in a custom-built furnace described in more details in reference [49]. The graphite carrier containing the sample was introduced into the preheated graphite reactor and annealed at 580 °C for 30 min in an Ar atmosphere with a background pressure of 300 mbar. In Figure S1 of the supporting information (SI) it is demonstrated that the XRD pattern of the NaInS$_2$ layer matches the database pattern for the compound with the *R-3mH* space group [50]. The Raman spectra of the NaInS$_2$ sample measured with 532, 633, and 785 nm excitation wavelengths are available in Figure S2.

**A CuIn$_5$S$_8$ thin-film reference sample** was prepared with an approach similar to NaInS$_2$. Cu and In were deposited on a Mo-coated high-strain glass (with low Na content) by co-evaporation. The composition of the precursor was verified by XRF to be [Cu]/[In] = 0.2. This metallic precursor was then sulfurized following the same procedure as described for NaInS$_2$ above. Figure S3 in the SI shows the XRD pattern of this sample, and it appears to be single-phase CuIn$_5$S$_8$. Figure S4 shows the multi-wavelength Raman spectra of the sample.

**An In$_2$S$_3$ thin-film reference sample** was prepared by annealing of a pellet of metallic In in sulfur following the same procedure as described above for the NaInS$_2$ sample.

## 2.2 Materials characterization

Scanning electron microscopy (SEM) was performed in a Zeiss LEO 1550 microscope. An acceleration voltage of 10 kV was used for all measurements. Raman spectroscopy was carried out at room temperature in a Renishaw inVia system using lasers with wavelengths of 532, 633, and 785 nm. Power densities in the range of 5−50 W/cm$^2$ were used. Grazing incidence (GI) and Bragg-Brentano (BB) XRD were performed with a Siemens D5000 system using Cu Kα radiation. All GIXRD measurements were performed with an incidence angle ($d_{inc}$) of 1°.

## 2.3 First-principles calculations

### 2.3.1 Computational setup

The first-principles calculations within density functional theory (DFT) were carried out using the Vienna *Ab initio* Simulation Package (VASP) employing projector augmented wave (PAW) method and Perdew-Burke-Ernzerhof (PBE) exchange-correlation functional. Pseudopotentials with the valence electron configurations of Cu 3d$^{10}$4s$^1$, In 5s$^2$5p$^1$, S 3s$^2$3p$^4$, and the cut-off energy of 350 eV were used. For the formation energy calculations, relatively dense Γ-centred Monkhorst-Pack grids with about 8000 k-points per reciprocal atom were adopted. Both lattice parameters and ionic positions were optimized when calculating formation energies until all ionic forces decreased below 10 meV/Å. The parameters used in the calculations for the Raman spectra simulations, when different from the already specified, are presented in section 2.3.6.

### 2.3.2 Data processing and visualization

Throughout the study, preparation of input files and analysis of results were greatly facilitated by the use of pymatgen (Python Materials Genomics) library [51], whereas the three-dimensional visualization of structures was done using the Visualization for Electronic and STructural Analysis (VESTA) software [52].

### 2.3.3 Polytype generation algorithm

To generate $CuInS_2$ polytypes, we developed a recursive algorithm filling all cationic sites in various supercells of the primitive zinc-blende (two atoms) or wurtzite (four atoms) cells with either Cu or In to reach the intended [Cu]/[In] = 1. All supercells containing up to 48 atoms for ZB-derived and 64 atoms for WZ-derived structures were considered. The starting set of inequivalent supercells was generated using the Alloy Theoretic Automated Toolkit (ATAT) [53] and then filling of the cationic sites was performed by a custom python script. As the goal was to produce structures obeying the octet rule, on-the-fly check function was incorporated to filter out all structures violating it. Moreover, the structures were constantly compared to the already identified ones and removed on-the-fly whenever duplication was discovered. There was no beforehand requirement to produce polytypes consisting of mixed CH- and CA-type domains only – such a motif was discovered later upon the inspection of generated dataset. The resulting dataset of structures after the ionic relaxation is accessible from the Materials Cloud repository [54] under the following identifier [55].

### 2.3.4 Assignment of CA-type fractions

The CA-type fraction was ascribed to every polytype based on the fraction of Cu atoms that have the same local coordination (within the second coordination sphere, i.e. with the nearest cations) as in the ordered $CuInS_2$ prototypes (i.e. ZB-CH, ZB-CA, WZ-CH, or WZ-CA).

### 2.3.5 Simulation of XRD patterns

The XRD patterns were simulated using diffraction module in pymatgen [51], which is based on the formalism laid out elsewhere [31]. X-ray wavelength of 1.5406 Å corresponding to Cu Kα was assumed. To add spectral broadening, Lorentzian smearing with the full width at half maximum (FWHM) of 0.2° was applied to all patters.

### 2.3.6 Simulation of Raman spectra

The off-resonance Raman spectra were simulated using "vasp_raman.py" utility written by Fonari and Stauffer [56], which in operation relies on the formalism described by Porezag and Pederson [57]. The method requires two ingredients: (1) frequencies of phonons at Γ-point and (2) derivatives of macroscopic dielectric tensor with respect to phonon vibrations for each mode. Both of these were computed within the density functional perturbation theory (DFPT) [58], as implemented in VASP package. The derivatives of dielectric tensor were computed numerically by displacing the atoms 0.01 Å in both positive and negative directions. Only the modes with wavenumbers above 100 $cm^{-1}$ were analysed. Instead of the PBE functional used for the formation energy analysis, PBE+U functional with the Hubbard U correction of 5 eV applied on Cu 3d was adopted for these calculations. The U correction was applied according to the formalism developed by Dudarev et al. [59]. The change in functional was necessitated by severe underestimation of the band gap energy by PBE, yielding inadequate Raman spectra. The magnitude of U correction energy was chosen based on the previous computational studies of $CuInSe_2$ [60]–[62], and it was found to deliver the intended band gap opening (although the band gap of 0.43 eV computed for *CH-CIS* after lattice relaxation was still grossly underestimated). Other Hubbard U values were tested and found to produce similar spectra. Prior to the phonon calculations, all structures were transformed into cells containing at least 14 atoms. To eliminate residual forces, atomic positions were re-optimized using a stricter atomic force threshold of 1 meV/Å. Based on our test, we found that experimental vibration frequencies are better reproduced when the lattice constants are fixed to their experimental values. As such, experimental lattices from Refs. [63], [64], [50], [65], [66] were set for secondary phases, whereas for all ZB- and WZ-derived structures the lattices of *CH-CIS* [67] and *WZ-CIS* [68] were used, respectively. The partially disordered $CuIn_5S_8$ (space group *F-43m*) [64] was converted into its ordered analogue by filling all mixed-cation sites by Cu and In in a checkerboard order. For the CIS polytypes, the lattice optimized with PBE were scaled by a constant such that the volumes of ordered *ZB-CH* and *WZ-CH* cells match the experimental values. The k-point density was changed to about 2500 points per reciprocal atom. For ordered CIS and secondary phases, the cut-off energy was set to 550 eV, whereas for polytypes this parameter remained equal 350 eV (the difference in Raman spectra was found to be negligible). The raw data files and optimized structures can be accessed under the following identifier [55]. To account for the spectral line broadening, all modes delivered by the vasp_raman.py script were smeared by the Lorentz function with FWHM of 10 $cm^{-1}$, unless otherwise specified.

## 2.4 Statistical analysis and raw data

The statistical analysis performed in the computational part of this study is limited to the construction of average XRD and Raman spectra for a collection of structures (ensembles) to describe cation-disordered *CIS* phases, as described below. The raw data and graphs for individual structures are accessible from the Materials Cloud archive [55]. The only pre-processing performed on the experimental data is normalization of all XRD and Raman measurements.

# 3. Results

### 3.1 *Ab initio* analysis of $CuInS_2$ polymorphs: Structures and energies

The possibility of forming different $CuInS_2$ polymorphs depending on the growth method is typically attributed to the small difference in their formation energies. This consideration is most often applied to *ZB-CA*, the energy of which is only about 2 meV/atom higher than of the ground state *ZB-CH* [19], [28]. Wurtzite $CuInS_2$ has also been found unstable with respect to *ZB-CH*, but the reported energies are surprisingly inconsistent [41], [69]. Still, Shen et al. [11] identified that *WZ-CH* has 1 meV/atom lower energy than *WZ-CA*. This agrees well with our calculations showing that metastable *ZB-CA*, *WZ-CH*, and *WZ-CA* have respectively 1.6, 5.4, and 6.2 meV/atom higher energy than the ground state *ZB-CH*.

The formation energies of disordered *d-ZB* and *d-WZ* phases are harder to assess. Experimentally, these are described as ideal lattices with partial cationic site occupancies. Computationally, the disorder is replicated by randomly filling the cationic sublattice with Cu and In in a sufficiently large supercell, yielding structures with large deviation from the octet rule. Several earlier reports have shown that presence of such deviations – (3In+Cu) and (3Cu+In) tetrahedra in the statistical mix – increases energy by 0.3-0.4 eV per unit [11], [18]. Due to the high energy cost, random distribution of cations should only be expected at high temperatures (we refer to it as the "high-temperature" model of disorder) and transform into a more stable mixture of ordered phases upon cooling [18]. When the disordered phases are grown at room or moderate temperature through colloidal synthesis, co-existing domains of ordered structures with zero deviation from the octet rule (and thus lower energy) are thought to form instead, as it was described for *WZ-CIS* nanoparticles before [11]. This scenario should be better represented by an ensemble of ordered low-energy polytypes with zero deviation from the octet rule. This type of disorder is referred to as the "room-temperature" model herein.

The computed energies and two representative polytypes are depicted in Figure 4. This figure compiles data for 269 ZB-based and 35 WZ-based polytypes, and it reveals linear correlations between the CuAu-type fraction and formation energies for both lattices. No dependence on the cell size was found and very little energy variation is seen for polytypes with the same CuAu-type fraction, meaning that they have the same probability of appearing in the material. This result means that, provided the stacking order does not violate the octet rule, interfacial energy between different polytypes is essentially zero. This result echoes the argumentation presented by Shen et al. for the interlaced *WZ-CIS* nanoparticles [11] and suggests that ZB-based *CIS* nanoparticles might have a similar compound morphology.

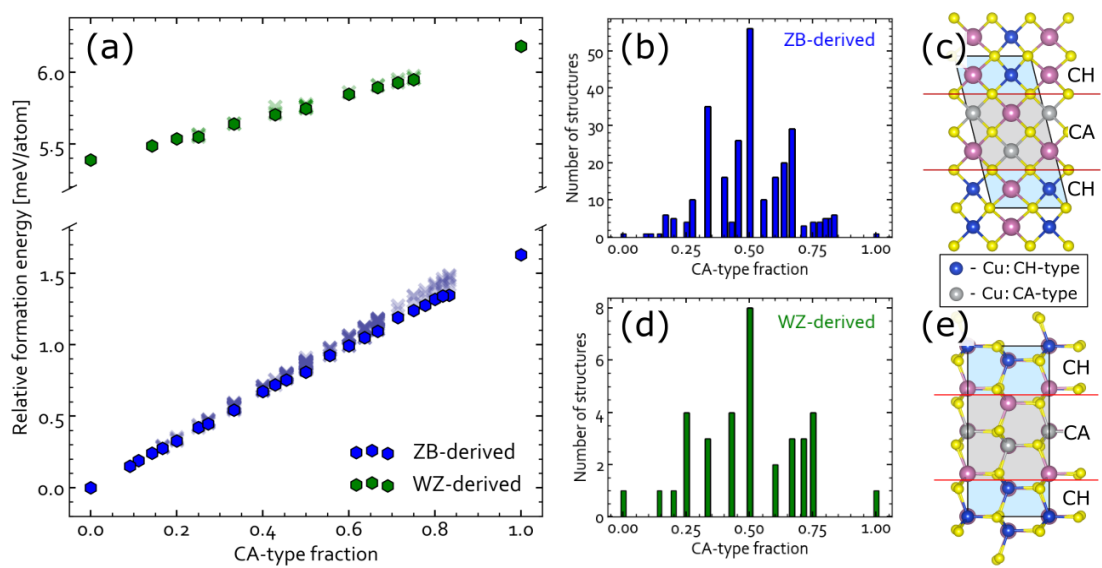

*Figure 4. Ground-state stability of CuInS$_2$ polytypes. (a) Formation energy of different CuInS$_2$ polytypes with respect to that of ZB-CH vs. the CA-type fraction. The hexagons represent the most stable structures for a given CA-type fraction; all other polytypes are represented by crosses. Distribution of (b) ZB- and (d) WZ-derived polytypes by the CA-type fraction. Examples of low-energy (c) ZB- and (e) WZ-derived polytypes with 50% CA-type fraction. Other polytypes can be accessed via the following identifier [55].*

To demonstrate the impact of such compound morphology on the material characterisation, we first simulated and then averaged XRD patterns for all polytypes with ZB- and WZ-derived structures. The results can therefore be treated as features of the metastable disordered phases of CIS. We find that this approach is similar to that developed by Jones and Stevanović for glassy solids [70], but more primitive as we ascribe all the structures identical statistical weights, which we think is a better approximation for the low-temperature disordered material. The obtained patterns are shown alongside those for the ordered and high-temperature disordered phases in Figure 5. As one can see, all unique reflections seen for the ordered phases vanish in the polytype-averaged patterns but resemble those for the high-temperature disordered *ZB-* and *WZ-CIS*. This similarity illustrates that the high- and room-temperature disordered materials are indistinguishable with XRD, explaining why no distinction between them has been conceived in experimental studies.

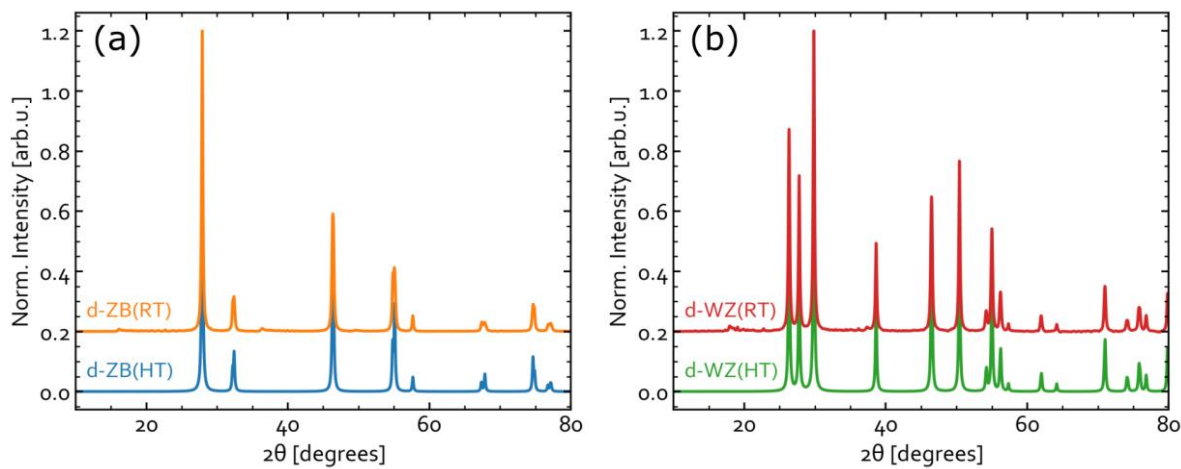

*Figure 5. Simulated XRD patterns for the fully cation-disordered (as seen at high temperature; HT) and ensemble-averaged (as expected at room-temperature; RT) representations of (a) d-ZB and (b) d-WZ phases. The d-ZB(RT) and d-WZ(RT) patterns were obtained by averaging the simulated patterns for 269 ZB-based and 35 WZ-based polytypes, whereas the d-ZB(HT) and d-WZ(HT) patterns are simulated by ascribing each cationic site the same partial occupancy of 50% In and 50% Cu. For all structures, the DFT-optimised geometries were scaled to match per atom volume of the experimental CH-CIS [67] or WZ-CIS structures [68].*

It is also interesting to point out that, accepting the linear correlation between formation energy and fraction of CA-type phase, both *ZB-CA* and *WZ-CA* must be unstable at all temperatures. This is because the configuration entropy, which contributes greatly to free energy at elevated temperatures, is higher for macrostates with a large number of microstates, peaking at 50% CA-type fraction for CIS, as hinted by Figure 4 (b),(d). Such macrostates should therefore stabilize upon heating. Since *ZB-CA* and *WZ-CA* are ordered, their configuration entropies are zero. As a result, any kinetic trapping necessary for the formation of metastable CA-type phases must be impossible. At the first glance, this conclusion contradicts numerous experimental observations, but the caveat is that *ZB-CA* in experiment has been characterised for samples produced by methods that can tip the energy balance in favour of *ZB-CA* (eg, possible substrate strains in epitaxial films [24]–[28], pressure-volume contributions in high-pressure synthesis [23]). At the same time, for the films grown by co-evaporation, the increased yield of *ZB-CA* in Cu-poor CIS might reflect the stabilization by off-stoichiometry [29], [30]. The discrepancy might also be rooted in the effects or parameters overlooked by our calculations (eg, vibration entropies). Herein, we do not attempt to identify the root cause but simply highlight the discrepancy for future studies.

## 3.2 Identification of secondary phases in Cu-poor CuInS$_2$ by Raman spectroscopy using reference materials and ab initio calculations

Raman spectroscopy is a powerful technique that allows identification of individual phases in a phase mixture with micrometre resolution. It has, for example, been successful in distinguishing phases in Cu$_2$ZnSnS$_4$-based materials that are not identifiable by XRD [71]. It can be useful to recognize contributions from various phases in CuInS$_2$-based absorbers, too. It is known that CIS grown under Cu-rich conditions results in formation of stoichiometric CIS and Cu$_x$S [12]. Since the Raman spectrum of Cu$_x$S is well known (main peak at 475 cm$^{-1}$ [72]), we herein focus on the phases that can form along with CIS when grown with a Cu-deficiency. The three phases found in the Cu-poor part of the phase diagram are $CuInS_2$, $CuIn_5S_8$, and $In_2S_3$. Even though $In_2S_3$ is not expected in any of the samples based on their stoichiometry, we include it here since the samples are grown under non-equilibrium conditions. In addition to the phases expected from the phase diagram, $NaInS_2$ was found to form due to the Na supply from the soda-lime glass. $NaInS_2$ has previously been observed in Cu-poor CIS prepared via reactive sputtering followed by annealing in $H_2S$ [5], and found to have Raman signatures at 158 and 289 cm$^{-1}$ [73]. The Raman spectrum of $CuIn_5S_8$ has also been documented in literature. The spectrum measured with an excitation wavelength of 514 nm was reported in 1991 by Gasanly *et al.* [74]. This data formed the basis for identification of $CuIn_5S_8$ in a phase mixture with CIS in several subsequent studies [34], [75]. The Raman spectrum of phase-pure $CuIn_5S_8$ has, however, not been reported with other excitation wavelengths, and is therefore worth examining. The Raman signature of $In_2S_3$ is better documented in literature (*e.g.* Refs. [76]–[78]). The spectra measured in this work are presented in Figure S5.

The Raman spectra of all reference materials measured with an excitation wavelength of 633 nm are compiled in Figure 6 (a). Spectra measured with 532 and 785 nm excitation are included in Figures S6 and S7 in the supporting information (SI).

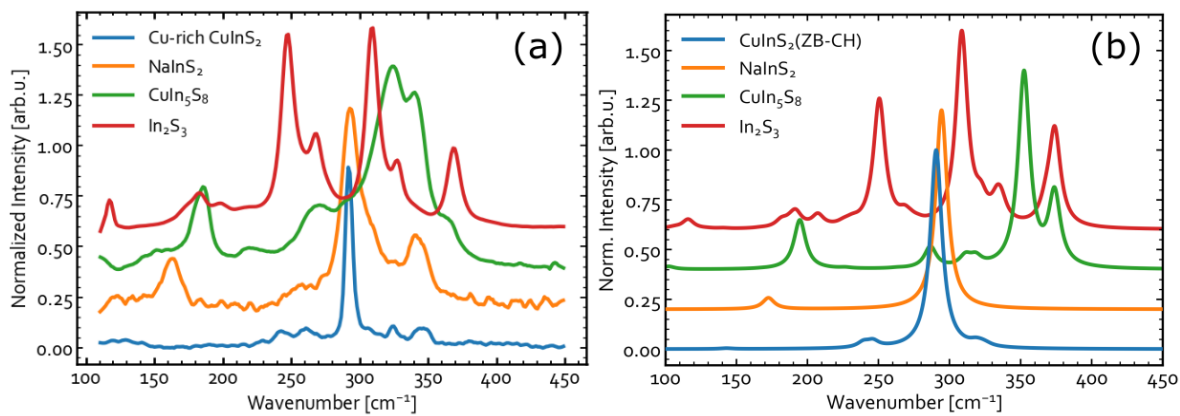

*Figure 6: Raman spectra of stable phases in the $CuInS_2$-$In_2S_3$ pseudo-binary system. (a) Experimental spectra recorded using 633 nm excitation laser. The $NaInS_2$ sample is measured with 5x higher excitation density than the other samples due to much weaker signal for this sample. (b) Simulated off-resonance Raman spectra for the same compounds.*

Included in Figure 6 (a) is the Raman spectrum of Cu-rich CIS ([Cu]/[In] = 1.39) in order to compare the reference samples to CIS free of phases that form under Cu-poor conditions. It is immediately clear that $In_2S_3$ is easily distinguishable from the other phases. The spectrum is dominated by peaks at 117, 138, 180, **247**, 267, **308**, 327, and 368 cm$^{-1}$ (bold font style indicates most prominent peaks), in good agreement with previous reports for β-$In_2S_3$ [76]–[78]. $CuIn_5S_8$ has peaks at **186**, 267, **324**, **340**, and 360 cm$^{-1}$, which is in agreement with the paper by Gasanly et al. [74], who observed a similar spectrum with relatively broad features and the most prominent peaks at 327, 341, and 360 cm$^{-1}$. The peak at 186 cm$^{-1}$ was less pronounced in the spectrum by Gasanly et al. This is likewise the case in our study when using the 532 nm excitation, as seen in Figure S4. The Raman spectrum of $NaInS_2$ is characterized by three main peaks at 163, **293**, and 341 cm$^{-1}$, while the $CuInS_2$ sample has peaks at 243, 260, **292**, 324, and 345 cm$^{-1}$, with the 292 cm$^{-1}$ mode clearly dominating the spectrum. This spectrum is typical for Cu-rich CIS measured under non-resonant conditions [75]. As one notices, the main mode of NaInS$_2$ is overlapping with the dominant signal for *CH-CIS*, which in the latter case is attributed to A$_1$ vibration mode of S-S bonds [79]. Thus, the 293 cm$^{-1}$ mode in NaInS$_2$ may have the same origin. It is noticed that the peaks in NaInS$_2$ are significantly wider than in CIS. This may be an indication of poorer crystal quality due to the non-optimized synthesis of NaInS$_2$. Based on the spectra presented in Figure 6 (a), it is clear that almost all the phases have unique peaks and may be distinguishable in Raman spectroscopy given that a large fraction of the phase is present in the probed volume. The main difficulty lies in the overlap of the main modes of *CH-CIS* and NaInS$_2$. Furthermore, the Raman response of NaInS$_2$ was found to be weak relative to CIS under the same measurement conditions. It is therefore unlikely to distinguish CIS and NaInS$_2$ using Raman in a phase mixture, unless it contains a very large NaInS$_2$ volume fraction.

To confirm the phase identification based on Raman measurements, off-resonant Raman spectra for different Cu-In-S phases were simulated. The computed results for *CH-CIS* phase of CuInS$_2$, thiospinel structures of In$_2$S$_3$ (space group *I4$_1$/amd*) [63] and CuIn$_5$S$_8$ (space group *F-43m*) [64], and delafossite NaInS$_2$ (space group *R-3m*) [50] are presented alongside the measured ones in Figure 6 (b). Clearly, all simulated spectra agree well with their experimental analogues, validating the assignment used in the previous section and proving that the employed methodology is indeed suitable for the material system of interest. Still, a small but systematic shift of all modes is observed for CuIn$_5$S$_8$, which could be because the partial cationic disorder seen experimentally [64] is not captured by our simulations. The key difference though is the absence of the experimental peak at 341 cm$^{-1}$ in the simulated Raman spectra of NaInS$_2$, suggesting that this sample might contain traces of yet another secondary phase. Based on the precursor film composition, this phase is expected to have [Na]/[In] ≈ 1.5, ie. Na-rich. To verify this hypothesis, additional calculations were performed for Na$_3$InS$_3$ (space group *C2/c*) [65] and Na$_5$InS$_4$ (space group *P2$_1$/m*) [66], as shown in Figure S2 (b). The obtained results show that the extra peak in the Raman spectra for NaInS$_2$ reference sample likely originates from Na$_3$InS$_3$.

### 3.3 Simulated Raman spectra of CuInS$_2$ polymorphs

In this section the simulated Raman spectra for different CuInS$_2$ polymorphs introduced earlier are presented. The results obtained for the four ordered and two disordered metastable phases are presented in Figure 7. The ordered structures here are modelled explicitly, whereas the disordered ones are represented by Raman spectra averaged for four (for ZB-derived) or three (for WZ-derived) different polytypes with 50% CuAu-type fraction. The individual Raman spectra for these and several other polytypes with different CA-type fractions are shown in Figure S8. As one can see, all six CIS phases have the dominant signals at the wavenumbers between 280 and 320 cm$^{-1}$. As expected, the position of the dominant peak for *ZB-CH* almost perfectly matches the experimental signal at 292 cm$^{-1}$ [75], but it unfortunately also overlaps with the dominant peak for *WZ-CH* CuInS$_2$ and even one intense peak for *WZ-CA*. Thus, *ZB-CH* and *WZ-CH* structures are indiscernible with Raman. It is expected, however, that this peak in most cases is dominated by *ZB-CH*, simply because it is the most stable at the conditions of growth and storage. The shoulder at around 305 cm$^{-1}$ in the experimental spectrum, which is typically attributed to *ZB-CA* in literature [39], can as well be explained by *d-ZB*, *d-WZ*, and *WZ-CA* as all these yield similar signals in the simulated spectra.

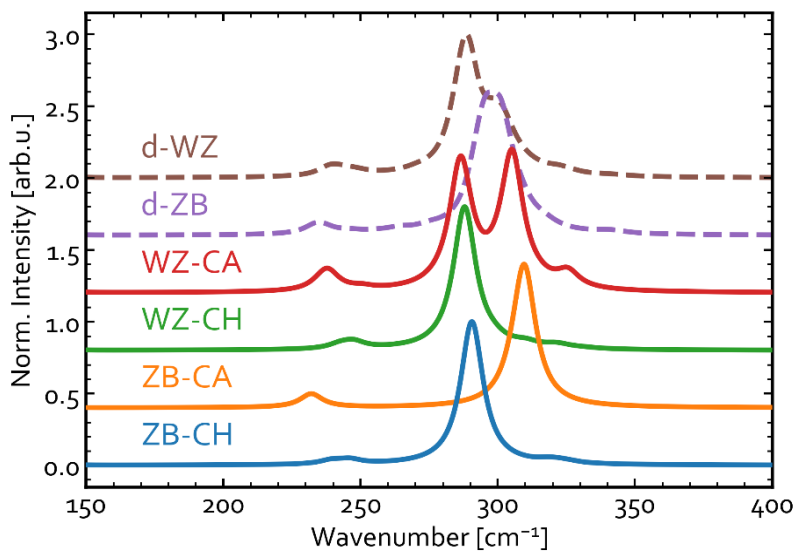

*Figure 7. Simulated Raman spectra for the different ordered structures and polytype-averaged disordered phases of CuInS$_2$.*

## 3.4 Properties of Cu-In-S precursors

While the previous sections focused on the results of calculations and measured Raman spectra of reference material, this section describes experimental work on sputtered CIS precursors with different compositions. The precursors sputtered without intentional substrate heating are characterized by Raman and GIXRD to understand their phase compositions. Figure 8 shows the diffractograms measured for four precursors with compositions varying from Cu-poor ([Cu]/[In] = 0.67) to Cu-rich ([Cu]/[In] = 1.39). The figure shows only a narrow range of 2θ angles for convenience of phase identification. The corresponding full-range XRD patterns are available in the SI (see Figure S9).

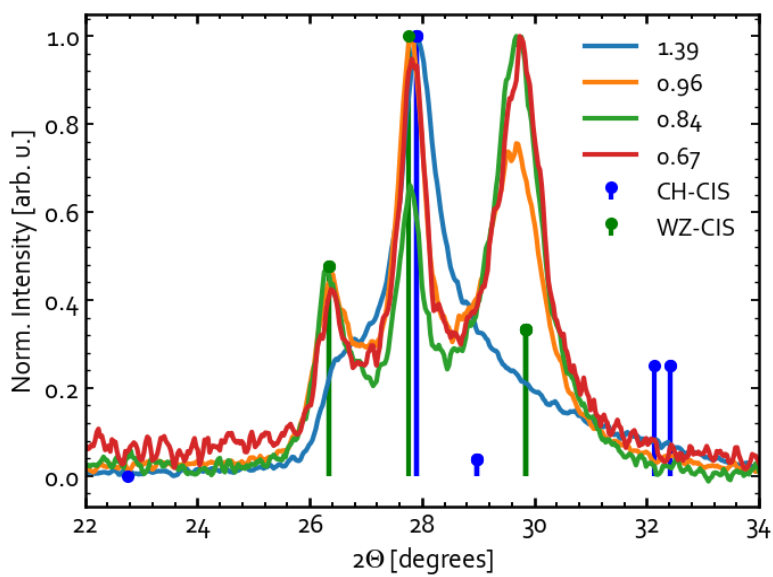

*Figure 8. GIXRD pattern of precursors with different [Cu]/[In] ratios expressed in the legend.*

The fact that clear peaks are present in the diffractograms indicates that the deposits partially crystallize during sputtering even without intentional substrate heating. A significant fraction of the layer may remain amorphous though. The reflections of the ICSD reference for the thermodynamically stable *CH-CIS* phase (collection #66865 [67]) are indicated with the blue markers in the figure. The diffractogram for Cu-rich precursor ([Cu]/[In] = 1.39) is found to be dominated by the reflections attributed to the *CH-CIS*, *CA-CIS*, and/or *SPH-CIS* phase. For the near-stoichiometric ([Cu]/[In] = 0.96) and Cu-poor precursors ([Cu]/[In] = 0.84 and 0.67), however, the thermodynamically stable chalcopyrite is clearly not the primary phase formed in the precursors. Instead, a significant contribution from *WZ-CIS* is observed, as indicated with green markers in Figure 8. The reference for the disordered $CuInS_2$ wurtzite structure (phase group *P63mc*) is the collection #163489 in the ICSD database [68]. The wurtzite phase has previously only been observed in CIS nanocrystals prepared by low-temperature solution-based processing [35], [36], [42], [68], [80], [81] and in electrodeposited films [32]. This is the first time the metastable phase has been found in sputtered $CuInS_2$ thin films, and it can be justified by the highly non-equilibrium growth process coupled with the low temperature causing the material to be trapped in the metastable phase. The observation that the *WZ-CIS* phase is more prominent in the Cu-poor precursors may indicate that wurtzite CIS is more tolerant to off-stoichiometry.

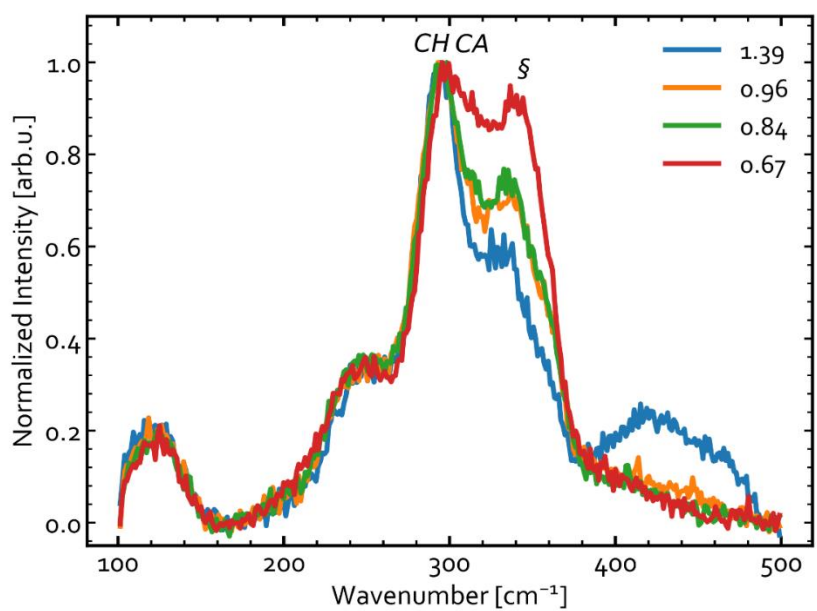

*Figure 9. Raman spectra measured with an excitation wavelength of 633 nm for the precursors with various chemical compositions expressed as [Cu]/[In] ratios in the legend.*

Figure 9 shows the Raman spectra of the precursors measured with an excitation wavelength of 633 nm. The spectra obtained with 523 and 785 nm excitation are shown in the SI (see Figures S14 and S15). The Raman spectra are dominated by relatively broad features, indicating small domain sizes. This is not surprising given the low processing temperature. In literature the main $A_1$ mode of *CH-CIS* is typically observed at around 292 cm$^{-1}$ [75], while the $A_1$ mode of *CA-CIS* appears at 305 cm$^{-1}$ [39]. It should, however, be kept in mind that the results of our calculations indicate that *d-ZB, d-WZ, WZ-CA,* and *ZB-CA* structures all have a peak near 305 cm$^{-1}$. The broad nature of the dominant peak could therefore include overlapping contributions from several polymorphs.

The main difference in the Raman spectra in Figure 9 relates to the mode labelled "§" centred around 340 cm$^{-1}$. This mode is clearly increased in intensity for samples with lower Cu content. The assignment of this peak is, however, not straightforward. The 340 cm$^{-1}$ vibration could coincide with E(L)/B$_2$(L) in *CH-CIS* at 345 cm$^{-1}$ [43]. The E(L)/B$_2$(L) typically does not show up in the spectrum of *CH-CIS* when using 532 or 633 nm excitation [82]. This mode can, however, be observed under resonant Raman conditions due to a break-down of the Raman selection rule as reported previously [75]. In our samples, a strong E(L)/B$_2$(L) mode is likewise observed when using 785 nm, which is a near-resonance conditions for CIS (see Figure S15 in the SI). The fact that a 340 cm$^{-1}$ mode also shows up under non-resonant conditions (633 nm excitation wavelength) indicates that either near-resonance is in fact playing a role or that this peak could relate to a different phase or compound than *CH-CIS*. The two candidates here are Na$_3$InS$_3$ and CuIn$_5$S$_8$ (see Figure 6 and Figure S2). CuIn$_5$S$_8$ is not found by XRD in the precursor, and presence of Na$_3$InS$_3$ would be surprising since a large Na surplus is required to form this compound. As demonstrated in the next section and in the SI (see Figure S24), a correlation exists between the quantity of *WZ-CIS* phase extracted from XRD and the 340 cm$^{-1}$ peak intensity in Raman with 633 nm excitation. The peak cannot, however, be assigned to any CIS polymorph under non-resonant conditions, as none of these have a strong signal around 340 cm$^{-1}$ in the simulation (see Figure 7). A peak at 340 cm$^{-1}$ was previously recorded for CuInS$_2$ nanoparticles [41], [83] and attributed to WZ-CA based on the analysis of computed vibration frequencies [41], but our results reveal that this mode is either very weak or silent under non-resonant conditions, suggesting that the signal could belong to hitherto unidentified phase or structure. All our characterization results discussed below seem to suggest that this phase has wurtzite lattice and [Cu]/[In] considerably below unity, which might be the stabilizing factor. An attempt was made to identify this structure by using high-throughput screening of CuInSe$_2$-In$_2$Se$_3$ pseudo-binary system, but all WZ structures discovered have energies high above the convex hull, even higher than ZB-derived ordered vacancy compounds (OVC) of the same composition, see Figure S25. Hence, the origin of the Raman signal at 340 cm$^{-1}$ remains to be determined.

The Raman spectra obtained using 532 nm excitation (see Figure S14 in SI) furthermore suggests that even the precursors with [Cu]/[In] = 0.84 contains inclusions of Cu$_x$S, as evidenced by peak at 475 cm$^{-1}$. The fact that this phase is observed in Cu-poor precursors further proves that the precursors are far from thermodynamic equilibrium.

## 3.5 Annealed absorbers
### 3.5.1 Microstructure

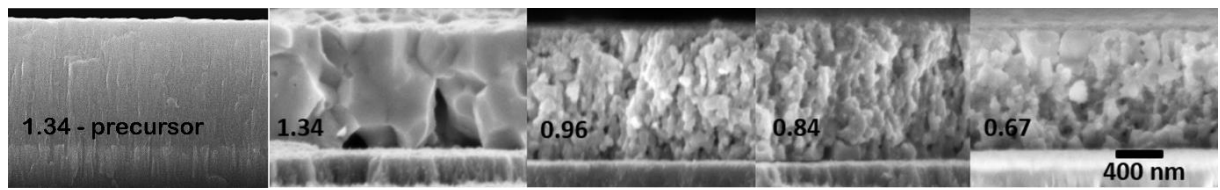

*Figure 10. SEM cross-section of a precursor and absorbers annealed at 600 °C for 30 min. The numbers in the images refers to the [Cu]/[In] ratios. The same scale bar applies to all images.*

Figure 10 shows SEM cross-sections of samples annealed at 600 °C for 30 min along with the Cu-rich precursor. It is observed that the composition of the precursors has a strong impact on the resulting microstructure after annealing. For the annealed Cu-rich sample ([Cu]/[In] = 1.34), large grains extending though the entire film thickness are formed. For all the Cu-poor samples, on the contrary, the layer consists of very fine grains. The most Cu-poor film stands out since a few larger grains are found at the surface. These are likely to be CuIn$_5$S$_8$ particles as demonstrated later by GIXRD and Raman spectroscopy. The observed microstructure is similar to co-evaporated CIS, where small grains are seen in material grown under Cu-poor deposition conditions and large grains form when growing Cu-rich absorbers [84]. The limited grain growth under Cu-poor conditions is a consequence of the narrow single-phase region of CIS, which restrains *CH-CIS* formation and stimulates decomposition of non-stoichiometric material into a phase mixture [14]. In order to determine the phase composition of the fine-grained layers, XRD and Raman measurements were performed on a series of samples annealed at different temperatures. An additional sample with 20 nm NaF deposited on the precursor surface prior to annealing was prepared and characterized to evaluate the impact of Na of the phase formation.

### 3.5.2 Analysis of co-existing phases

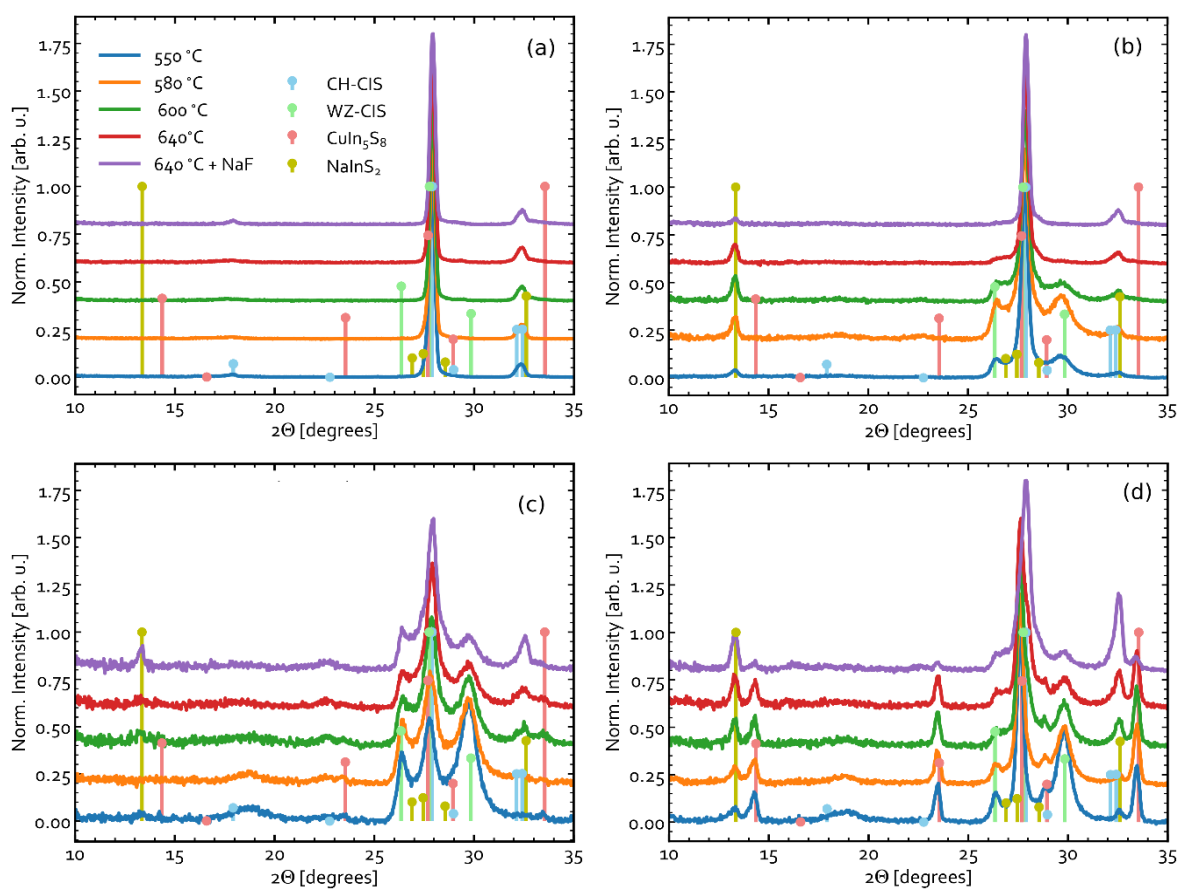

*Figure 11: GIXRD ($d_{inc} = 1°$) patterns of Cu-In-S samples after annealing at various temperatures. The [Cu]/[In] ratios of the samples are (a) 1.39, (b) 0.94, (c) 0.84, and (d) 0.67. The reference patterns in this figure are CH-CIS(#66865) [67], WZ-CIS(#163489)[68], CuIn$_5$S$_8$ (#16423)[64], and NaInS$_2$ (#640036)[50], as taken from the ICSD FIZ Karlsruhe database. The same legend applies to all subfigures.*

Figure 11 shows a magnified view of the XRD patterns measured for the annealed Cu-In-S samples. This range contains signatures of the relevant phases. The complete XRD patterns are available in the SI (see Figures S10-S13). As seen in Figure 11 (a), the Cu-rich samples do not contain significant amounts of *WZ-CIS* after annealing irrespective of the temperature. The small *WZ-CIS* contribution seen in the precursor is gone after annealing, and the pattern is now dominated by *CH-CIS* as evidenced by the peak at $2\theta \approx 17.9°$. The presence of Cu$_x$S, which could be expected at such Cu-rich compositions, was not detected by XRD (see Figure S10 in the SI), presumably due to poor crystallinity. A signal corresponding to Cu$_x$S phase was instead recorded by Raman spectroscopy (see Figure S16). For the Cu-rich samples, the main impact of annealing temperature is sharpening of the XRD reflections due to the promoted grain growth at higher temperatures. In the Cu-poor samples, the co-existence of *CH-CIS, CA-CIS,* or *SPH-CIS* and *WZ-CIS* polymorphs persists after annealing. The near-stoichiometric sample ([Cu]/[In] = 0.96) in Figure 11 (b) contains a significant *WZ-CIS* fraction after annealing at the lowest temperatures, but its amount is gradually reduced with temperature. For the precursor with an additional NaF layer, the *WZ-CIS* phase is almost vanished after annealing, and the XRD pattern becomes similar to that of the Cu-rich sample. The unique peaks of *CH-CIS* are however not clearly visible, making it impossible to rule out that the dominating phase is *SPH-CIS*. In the more Cu-poor sample ([Cu]/[In] = 0.84), the *WZ-CIS* phase fraction is also reduced with increasing annealing temperature, but a substantial quantity remains even at the highest annealing temperature (see Figure 11 (c)).

Surprisingly, the sample with [Cu]/[In] = 0.84 (see Figure 11 (c)) does not appear to contain $CuIn_5S_8$ phase. Based on the $Cu_2S$-$In_2S_3$ phase diagram, a volume fraction of 9% $CuIn_5S_8$ is expected for this composition. The excess In is instead incorporated into $NaInS_2$ as evidenced by the XRD peaks at 13.2º and 32.4º that emerged for all Cu-poor samples (see Figure 11(b)-(d)). This is despite no NaF layer was deposited and the Na must have in-diffused form the glass substrate. Therefore, it can be concluded that, provided sufficient Na is available, $NaInS_2$ is more likely than $CuIn_5S_8$ to form and accommodate excess In not incorporated into $CuInS_2$.

In the most Cu-poor sample with [Cu]/[In] = 0.67 (see Figure 11 (d)), the $CuIn_5S_8$ phase is clearly observed after annealing. A volume fraction of about 28% $CuIn_5S_8$ is expected if the *CH-CIS* and $CuIn_5S_8$ phases were both stoichiometric and in thermodynamic equilibrium. The quantity of $CuIn_5S_8$ may, however, be reduced due to formation of $NaInS_2$, similarly to the sample with [Cu]/[In] = 0.84. This trend not only confirms the active role of Na for crystallization of CIS absorbers, but also highlights the competition between different phases for the excess In.

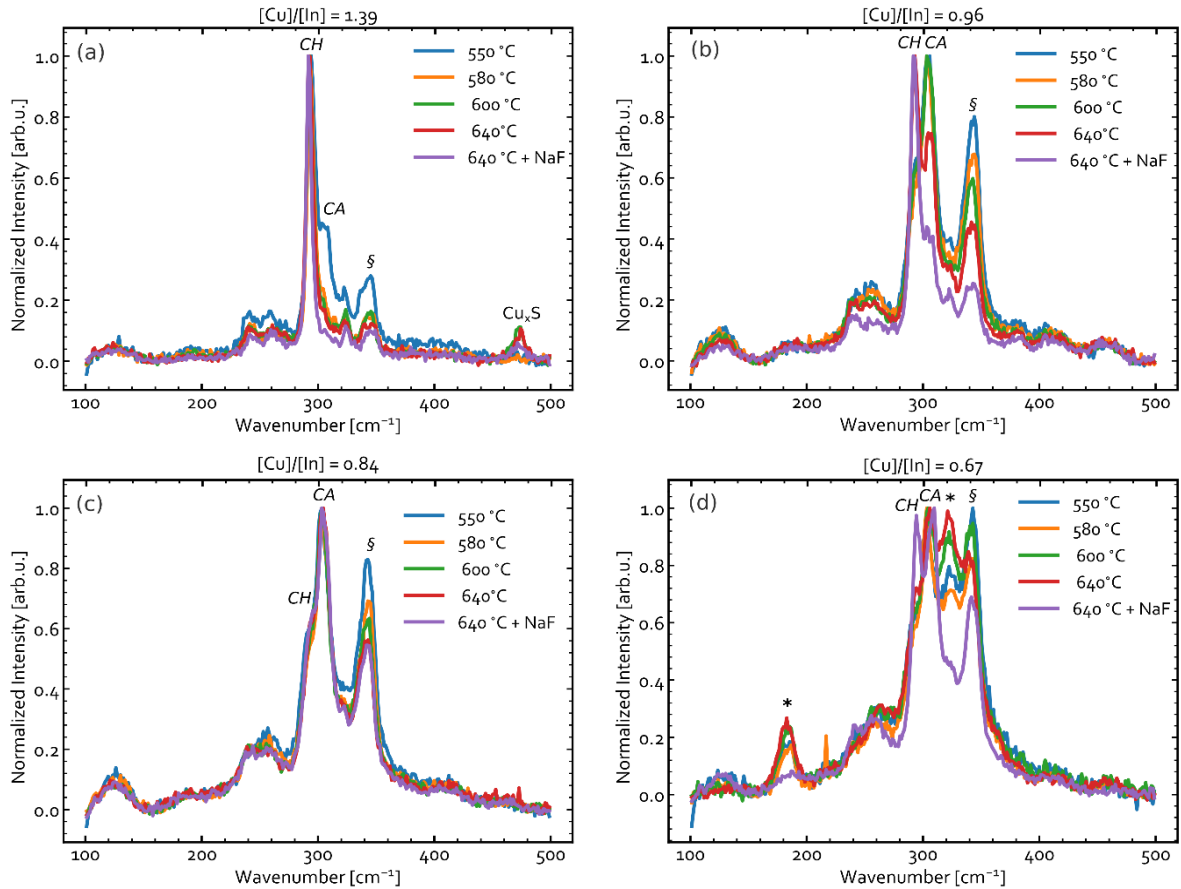

*Figure 12: Raman spectra measured with an excitation wavelength $\lambda_{exc}$ = 633 nm for the annealed Cu-In-S thin films with the [Cu]/[In] ratios of (a) 1.39, (b) 0.94, (c) 0.84, and (d) 0.67. The main contributions from the CIS polymorphs are labelled with "CH" and "CA". The "*" symbol marks the peaks attributed to $CuIn_5S_8$, while "§" indicates the ambiguous mode at 340 $cm^{-1}$.*

Figure 12 shows Raman spectra measured for the annealed samples with an excitation wavelength of 633 nm. The corresponding spectra obtained using 532 and 785 nm excitation are available in the SI (see Figures S16-S23). The Cu-rich sample (see Figure 12 (a)) has a noticeable contribution from $Cu_xS$ at 475 $cm^{-1}$. This mode is amplified under the 532 nm laser, as seen in Figure S16. When the samples are annealed at the lowest temperature of 550 °C, a contribution of *CA-CIS* at 305 $cm^{-1}$ is visible in the spectrum and the peak at 340 $cm^{-1}$ (marked "§") is prominent. For samples annealed at higher temperatures, these contributions are reduced and the spectrum is dominated by the $A_1$ of *CH-CIS* at 292 $cm^{-1}$. Here it must be kept in mind that the mode labelled "*CH*" could also relate to other CIS structures like WZ-CH or WZ-CA as shown above (see Figure 7). In the Cu-poor samples (see Figure 12(b)-(d)), the *CA-CIS* mode at 305 $cm^{-1}$ is the most prominent. Although *CA-CIS* has been claimed observable by XRD, it was not possible to ascribe any reflections to this phase in our samples. The presence of *CA-CIS* is therefore only supported by the Raman measurements, and may be questioned in light of the predicted peak overlap with other CIS structures. For the samples with [Cu]/[In] = 0.96, the *CA-CIS* signal dwindle compared to the *CH-CIS* $A_1$ mode with annealing temperature and addition of NaF. Curiously, this is not the case for the sample with [Cu]/[In] = 0.84. At the same time, the ambiguous mode at about 340 $cm^{-1}$ diminish with temperature in all samples (see Figure 12).

Raman spectra of the most Cu-poor samples ([Cu]/[In] = 0.67 in Figure 12 (d)) contain peaks at 186 and 324 cm$^{-1}$. These modes were not recorded for any other sample and are assigned to CuIn$_5$S$_8$ (see Figure 6). In agreement with XRD, this phase is only evident in the most Cu-poor sample, and strongly reduced when NaF has been added to the precursors.

## 4. Discussion

The formation of *CH-CIS*, *WZ-CIS*, *CA-CIS*, CuIn$_5$S$_8$, and NaInS$_2$ in different proportions is detected in Cu-poor CIS samples. All these phases except for *CA-CIS* are observed by XRD. *CA-CIS* is only detected by Raman spectroscopy and may relate to either ZB-CA, WZ-CA or disordered CIS phases (see Figure 7). NaInS$_2$ is only observed in XRD (see Figure 11). Based on the trends in the Raman spectra, it is clear that the *CA-CIS* peak (305 cm$^{-1}$) is of a different origin than *WZ-CIS* and CuIn$_5$S$_8$. *CA-CIS* is not eliminated upon heating the sample with [Cu]/[In] = 0.84, unlike the 340 cm$^{-1}$ mode that remained unassigned. It is therefore proposed that at least three CIS polymorphs co-exists in the Cu-poor CIS: *CH-CIS, WZ-CIS,* and *CA-CIS*. It may not be excluded that a disordered phase like *SPH-CIS* is present in the samples as well, even though it is unlikely at room temperature as explained above. The co-existence of interlaced *CH-* and *CA*-ordered CIS domains, as previously observed in nanoparticles [11], might be a more accurate description of the sample morphology at nanoscale. The relative contribution of *WZ-CIS* always decreases with increasing temperature, but the Raman peak related to *CA-CIS* is more persistent in the Cu-poor material. This likely stems from the fact that the formation energy of *CA-CIS* is much closer to *CH-CIS* than *WZ-CIS* (see Figure 4).

One can speculate about the causes behind the formation of large fractions of the metastable phases in the Cu-poor samples. In the more explored selenide Cu-In-Se system, Boehnke and Köhn found that *SPH* phase is stabilized down to room temperature when the composition falls in the range 0.32 < [Cu]/[In] < 0.49 and 1.13 < [Se]/([In]+[Cu]) < 1.20 [13]. The idea that the metastable CIS polymorphs can be stabilized by off-stoichiometry is further supported by the phase diagram published by Binsma et al. [12], who found that the *WZ-CIS* and *SPH-CIS* modifications persist at up to 100 °C lower temperatures when grown with either excess Cu$_2$S or excess In$_2$S$_3$ (see Figure 1). It has furthermore been argued that the disordered nature of *WZ-CIS* and *SPH-CIS* phases with random distribution of Cu and In on the cation sublattice enhance their flexibility to off-stoichiometry [34]. It seems plausible that the formation of metastable phases is triggered by off-stoichiometry of the precursors, not only with regard to [Cu]/[In] but also excess S which shifts the composition away from the Cu$_2$S-In$_2$S$_3$ tie-line (provided behaviour of Cu-In-S and Cu-In-Se systems are similar). Another potential explanation is the shift in phase balance caused by S deficiency in the films. We consider this scenario unlikely, but it cannot be ruled out since S powder was gone by the end of sulfurization, and hence, saturation of the films with gaseous S might have been incomplete (or even reversed by S loss during the cooldown).

It is more surprising that the metastable *WZ-CIS* phase persists after annealing at 640 °C. A previous study on CIS nanoparticles found that *WZ-CIS* and *CH-CIS* can co-exist after annealing at temperatures between 200 and 400 °C [37]. When annealing the nanoparticles at 500 °C, however, the wurtzite phase completely transformed into *CH-CIS* [37]. Another study found that *WZ-CIS* converted partially into *CH-CIS* upon heating at 406 °C, but that heating to 600 °C was necessary to eliminate all traces of this phase completely [68]. It is therefore obvious that *WZ-CIS* persists at even higher temperature in this study. This may relate to the overall Cu-poor compositions of the films, in which the *WZ-CIS* precipitates are presumably off-stoichiometric while *CH-CIS* phase is unable to accommodate high Cu deficiency. As such, converting Cu-poor *WZ-CIS* into the thermodynamically stable structure would necessitate phase separation into *CH-CIS* and $CuIn_5S_8$, which might be slow due to the high reaction barrier.

From the results presented here follows that deposition of high-quality single-phase Cu-poor $CuInS_2$ is not a trivial task. It could be accomplished by keeping stoichiometry within the narrow single-phase region, but the approach of low-temperature sputtering followed by annealing is particularly unsuited for the undertaking. The situation is exacerbated by the use of binary sputter targets in this work. The complex interplay between the metastable CIS phases in addition to the competition between $CuIn_5S_8$ and $NaInS_2$ for the excess In are complicating the matter even further. And once the metastable *WZ-CIS* is formed due to the off-stoichiometry or otherwise, it is extremely difficult to eliminate by post-deposition annealing. Hence, an alternative deposition strategy with more precise composition control or a different post-processing route may need to be employed. For instance, sputtering from metal targets would certainly not produce the metastable phases in the first place, which might be the reason why the best $Cu(In,Ga)S_2$ solar cell to date was fabricated with the absorber grown by sulfurization of sputtered metallic precursors in $H_2S$ [8]. However, the narrow single-phase region of *CH-CIS* may still cause issues during the subsequent sulfurization.

The adoption of co-evaporation strategy, which has been particularly successful for selenide $Cu(In,Ga)Se_2$, might also be a viable alternative. Since this route ensures sulfur overpressure and the growth process is closer to thermodynamic equilibrium, the formation of metastable phases might be supressed and their impact on devices alleviated. Indeed, three-stage co-evaporation has been recently demonstrated to yield devices with 14-15.5% efficiency [9], [85]. At the same time, Thomere *et al.* described an unexpected bilayer morphology of the absorbers stemming from segregation of spinel $CuIn_5S_8$-like and Cu-poor tetragonal $Cu(In,Ga)S_2$ phases during the second deposition stage [86]. The root cause of this behaviour is the narrow single-phase region of *CH-CIS*, which is further complicated by the introduction of gallium. On the contrary, Shukla *et al.* did not report the bi-layer morphology for films produced via a similar three-stage co-evaporation approach [9], indicating that the phase formation is strongly dependent on the deposition parameters and that the described issues can be circumvented, in principle.

## 5. Conclusions

Several metastable $CuInS_2$ phases were observed with a series of XRD and Raman measurements on Cu-poor films produced by room temperature sputtering from binary targets followed by a high-temperature annealing. Most of these phases have been reported earlier and are already ascribed distinct XRD reflections and/or vibration modes, but our simulations based on the first-principles calculations reveal that this assignment is ambiguous. Most notably, the $A_1$-S-S vibration mode at 305 $cm^{-1}$ in Raman spectra of $CuInS_2$, which is typically attributed to *CA-CIS,* can as well be ascribed to disordered *SPH-CIS* or *WZ-CIS*, as well as ordered WZ-CH or WZ-CA structures. Previously suggested assignment of the Raman peak at about 340 cm$^{-1}$ to *WZ-CIS* phase is also shown to lack sufficient evidence. Based on the analysis of computed formation energies for hundreds of CIS polytypes and existing literature, we conclude that CIS grown via non-equilibrium routes is likely to exhibit complex morphology of interlaced domains with closely-related ordered structures. Still, the metastable phases with different lattice types could be distinguished by XRD, which allowed us to detect wurtzite $CuInS_2$ for the first time in thin films produced by sputtering. While the presence of metastable phases is not particularly surprising in the precursors, the persistence of *WZ-CIS* after annealing at 640 °C is unexpected and could pose a practical challenge. The co-existence and complex interplay between *WZ-CIS*, *CH-CIS*, *CA-CIS*, $CuIn_5S_8$, and $NaInS_2$ in the Cu-poor CIS precludes this approach of producing homogeneous Cu-poor CIS for solar cell applications. It is argued that the metastable phases are stabilized by off-stoichiometry at the precursor deposition stage and remain kinetically trapped during post-deposition annealing. Future attempts to make Cu-poor CIS must therefore focus on discovering alternative strategies for achieving perfect stoichiometry in order to avoid or effectively eliminate the undesired phases.

# 6. Supporting Information

Supporting Information is available from the Wiley Online Library or from the author.

# 7. Acknowledgements

Funding from the Swedish Research Council (2019-04793) and the Swedish Foundation for Strategic Research (grant no. RMA15-0030) is gratefully acknowledged. The computations and data handling were enabled by resources provided by the Swedish National Infrastructure for Computing (SNIC), partially funded by the Swedish Research Council through grant agreement no. 2018-05973.

*Supporting Information for*

# Experimental and theoretical study of stable and metastable phases in sputtered CuInS$_2$

J. K. Larsen[1,§,*], K. V. Sopiha[1,§,*], C. Persson[2,3], C. Platzer-Björkman[1], M. Edoff[1]

[1] Division of Solar Cell Technology, Department of Materials Science and Engineering, Uppsala University, Box 534, SE-75237 Uppsala, Sweden

[2] Centre for Materials Science and Nanotechnology/Department of Physics, University of Oslo, Blindern, Box 1048, NO-0316 Oslo, Norway

[3] Department of Materials Science and Engineering, Royal Institute of Technology, SE-10044 Stockholm, Sweden

\* Corresponding authors: J. K. Larsen (jes.larsen@angstrom.uu.se), K. V. Sopiha (kostiantyn.sopiha@angstrom.uu.se)

§ These authors contributed equally to the work

# 1. Reference materials
## 1.1. NaInS$_2$ reference sample

A NaInS$_2$ reference sample was produced in order to investigate the characteristic Raman peaks of the material. The samples were produced by deposition of 50 nm NaF by e-beam evaporation on soda-lime glass, 30 nm In was deposited onto the NaF. The film was subsequently annealed at 580 °C for 30 min in a graphite reactor with 20 g of elemental sulfur and a 300 mbar Ar background pressure.

The thin film was characterized by GIXRD and Bragg-Brentano XRD to verify that NaInS$_2$ had formed as shown in Figure S 1. Based on the reference pattern for NaInS$_2$ with card #640036 in the ICSD database [1], it is concluded that the NaInS$_2$ phase was successfully produced with this approach.

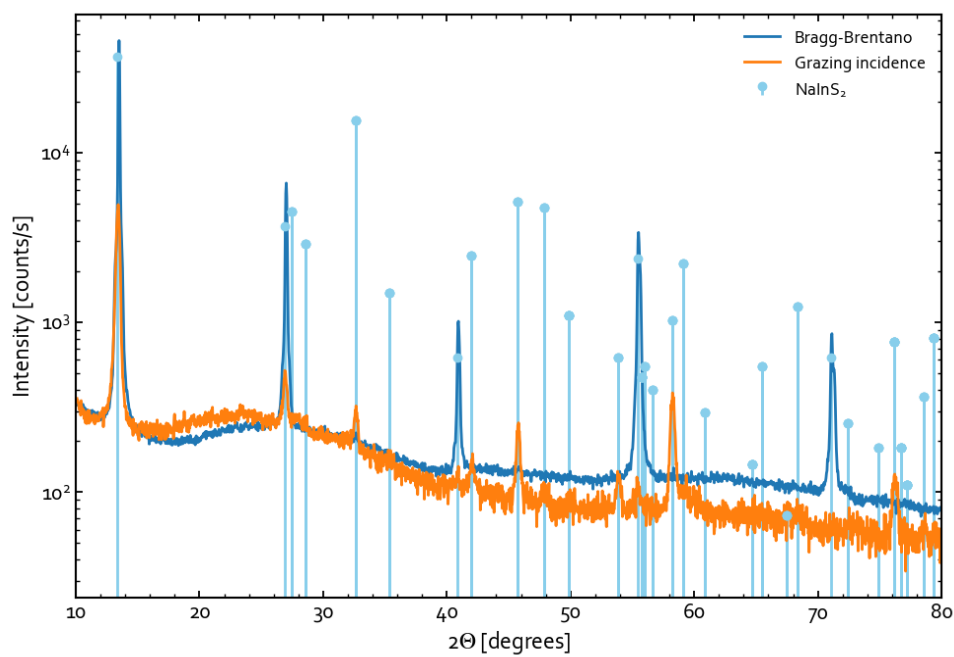

*Figure S 1: GIXRD and Bragg-Brentano XRD patterns of a NaInS$_2$ thin film on soda-lime glass. The reference pattern for NaInS$_2$ with card no 640036 from the ICSD database is included to indicate the expected reflections of the compound [1].*

**Error! Reference source not found.** Figure S2 shows the Raman spectra of NaInS$_2$ measured with different excitation wavelengths. Akin to CuInS$_2$, it is noticed that the modes at 258 cm$^{-1}$ and 341 cm$^{-1}$ are enhanced when using the 785 nm laser. This is somewhat surprising since it would not be expected that NaInS$_2$ in resonance with the 785 nm laser. The band gap of NaInS$_2$ has been reported to be around 2.3 eV [2]–[4]. The compound is therefore expected to be in resonance with the 352 nm laser.

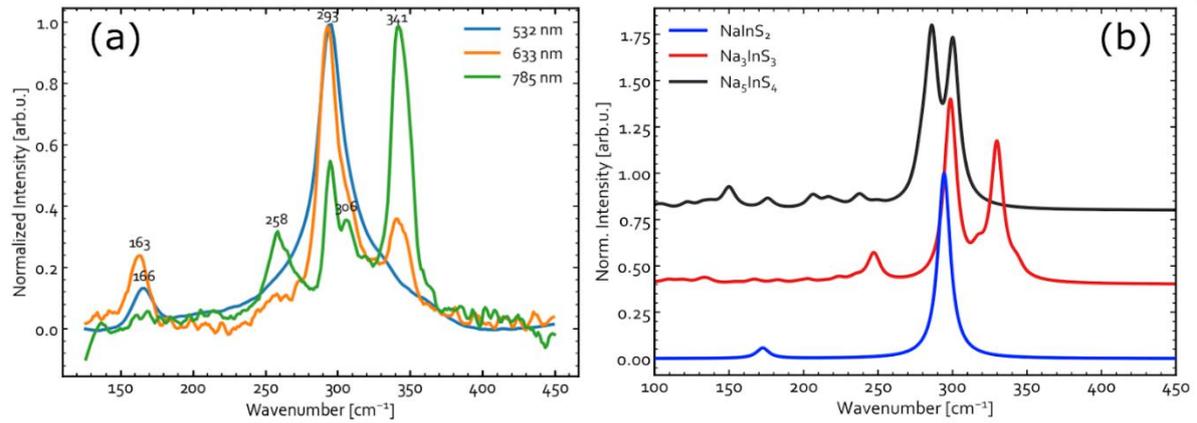

*Figure S 2: Raman data for Na-In-S phases. (a) Experimental spectra of NaInS$_2$ reference sample measured using different excitation wavelengths. (b) Simulated spectra of different Na-In-S phases with [Na]/[In]≥1.*

## 1.2. CuIn$_5$S$_8$ thin film reference

A CuIn$_5$S$_8$ reference sample was produced to unambiguously verify the Raman spectrum of the compound. Cu and In was deposited on high strain, low Na glass in the ratio [Cu]/[In] = 0.2 by co-evaporation. The precursor was then annealed at 580 °C in a sulfur-containing atmosphere to produce CuIn$_5$S$_8$. Based on the XRD pattern in reference [5], it can be confirmed that the CuIn$_5$S$_8$ was formed (see Figure S 3). Since no unidentified peaks are seen in the pattern, it appears that the sample consists of single phase CuIn$_5$S$_8$. Raman spectra of the sample measured with various excitation wavelengths is available in Figure S 4.

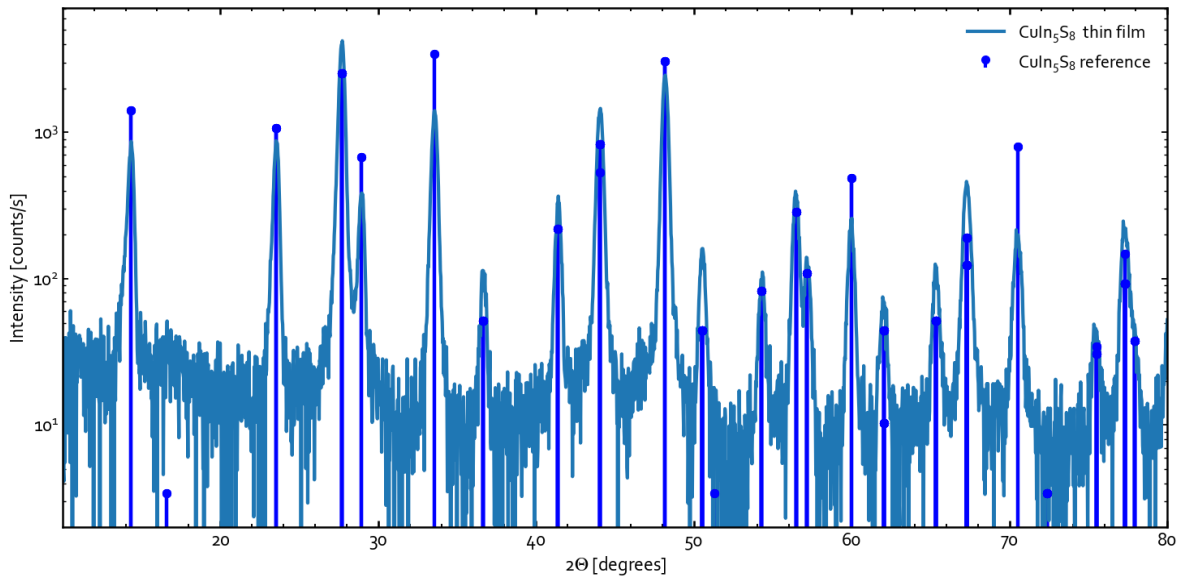

*Figure S 3: GIXRD ($d_{inc}$=1°) of CuIn$_5$S$_8$ thin film. The reference pattern in is card #16423 in the ICSD, originally published in reference [5].*

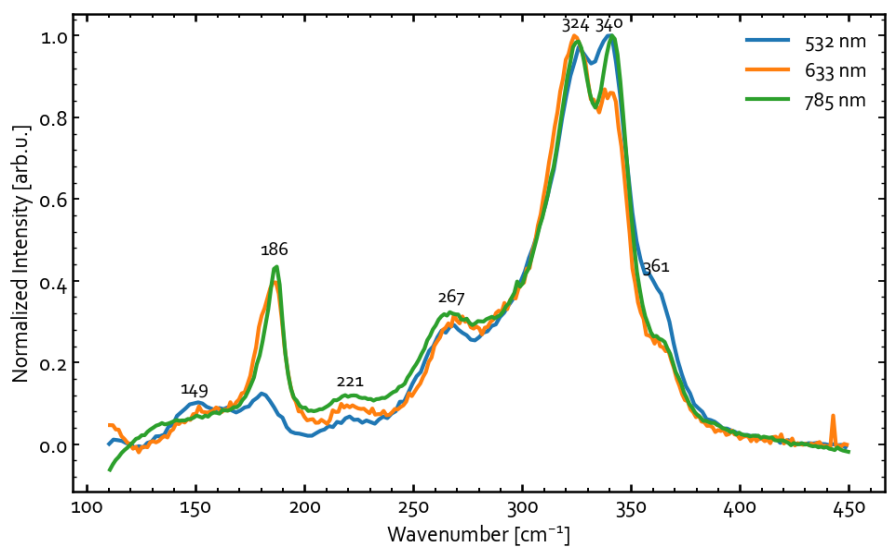

*Figure S 4: Raman spectra of CuIn$_5$S$_8$ measured with various excitation wavelengths indicated in the legend.*

### 1.3. In$_2$S$_3$ reference

In$_2$S$_3$ was produced by sulfurization of a Indium metal shot at 580 °C. The Raman spectrum measured with 532, 633, and 785 nm excitation is shown in Figure S 5. The characteristic peaks observed in the material are in good agreement with previously published Raman spectra of In$_2$S$_3$ [6]–[8].

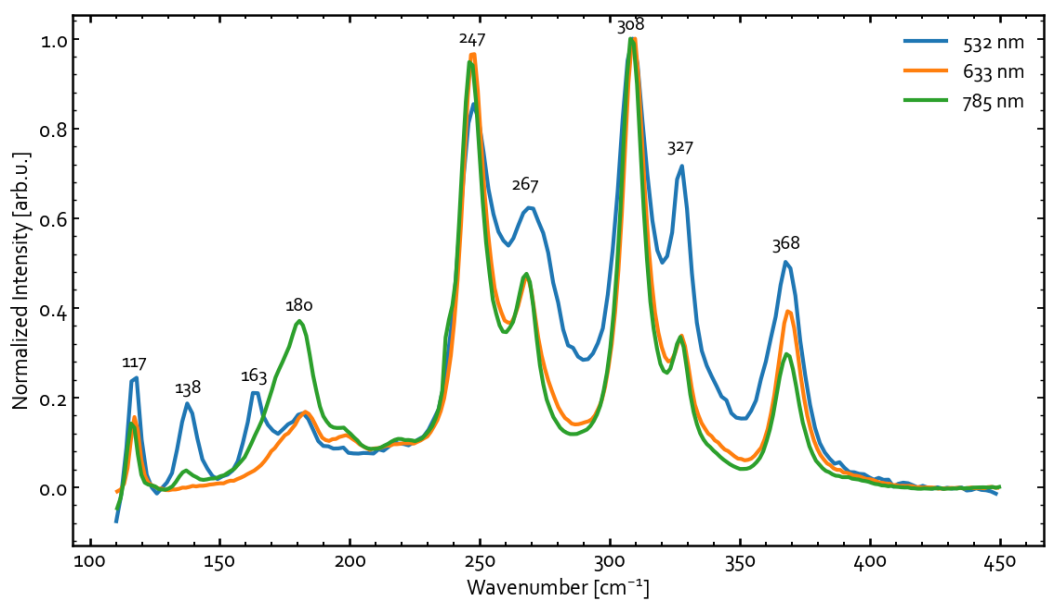

*Figure S 5: Raman spectra of In$_2$S$_3$ measured with various excitation wavelengths indicated in the legend.*

### 1.4. Comparative Raman spectra

In the main paper the Raman spectra measured with an excitation wavelength of 633 nm were presented. Since the spectra are sensitive to the excitation wavelength we supplement with measurements performed with $\lambda_{exc}$ = 532, 785 nm in Figure S 6 and Figure S 7.

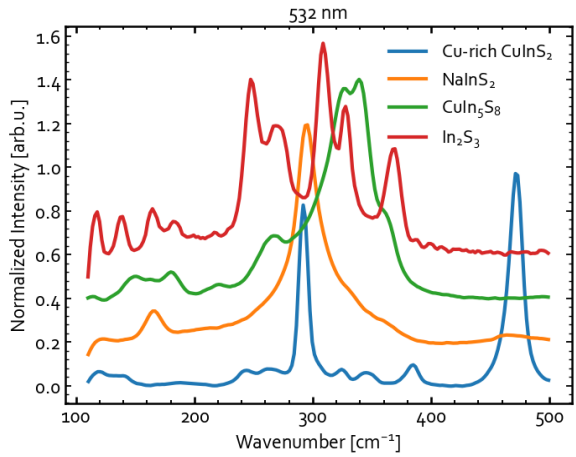

*Figure S 6: Raman spectra of reference material measured with $\lambda_{exc}$ = 532 nm. The peak in the CuInS2 sample at 475 cm$^{-1}$ belongs to Cu$_2$S.*

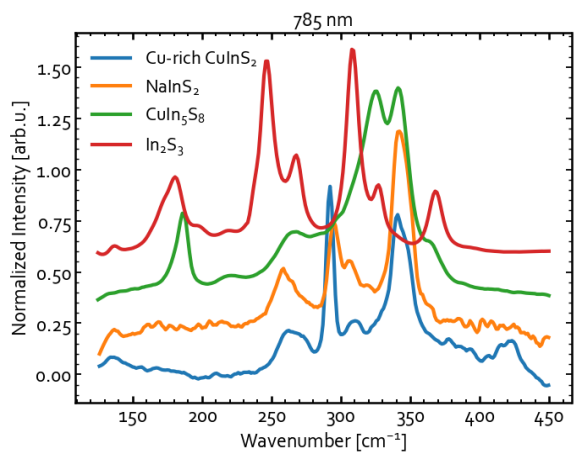

*Figure S 7: Raman spectra of reference material measured with $\lambda_{exc}$ = 785 nm.*

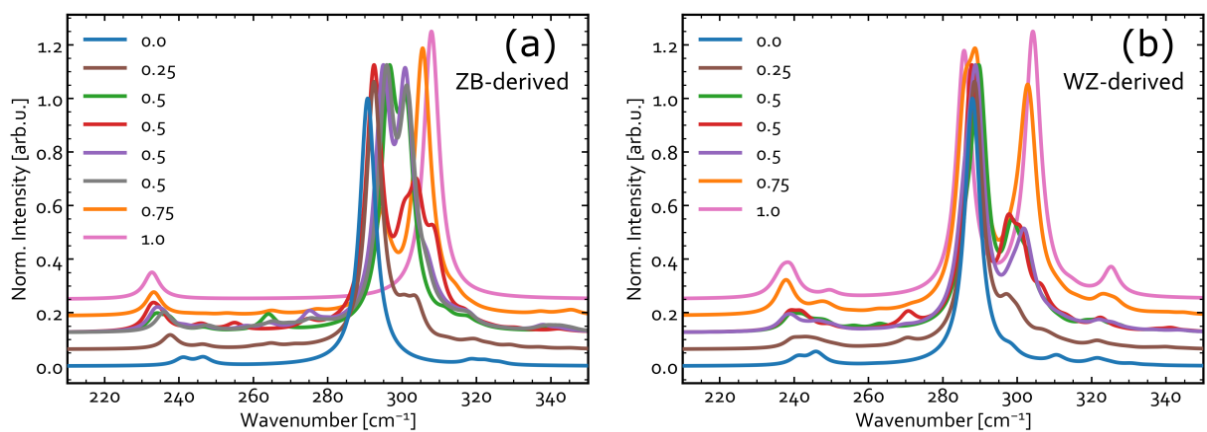

*Figure S 8: Simulated Raman spectra of CIS polytypes with (a) ZB-derived and (b) WZ-derived structures. The numbers in legends represent CA-type fractions. To better resolve the peaks, Lorentzian smearing with a smaller FWHM of 5 cm$^{-1}$ is used for these spectra. The red curve in (a) is notably different from the Raman spectra of other CIS polymorphs with CA-type fraction of 0.5 – the outlier is a polytype consisting of two large domains (CH-type and CA-type of 16 atoms each), whereas all other ZB-derived polytypes are composed of multiple small intermixed domains. This result indicates that further increase in the domain size may result in Raman spectra consisting of mixed CH- and CA-type features, but this assumption requires further analysis.*

## 2. XRD patterns with wider ranges

In the main paper a limited 2θ range of the GIXRD measurements was presented in order to more easily distinguish key features. For completeness the full range of the measurements are shown in Figures S9 – S13.

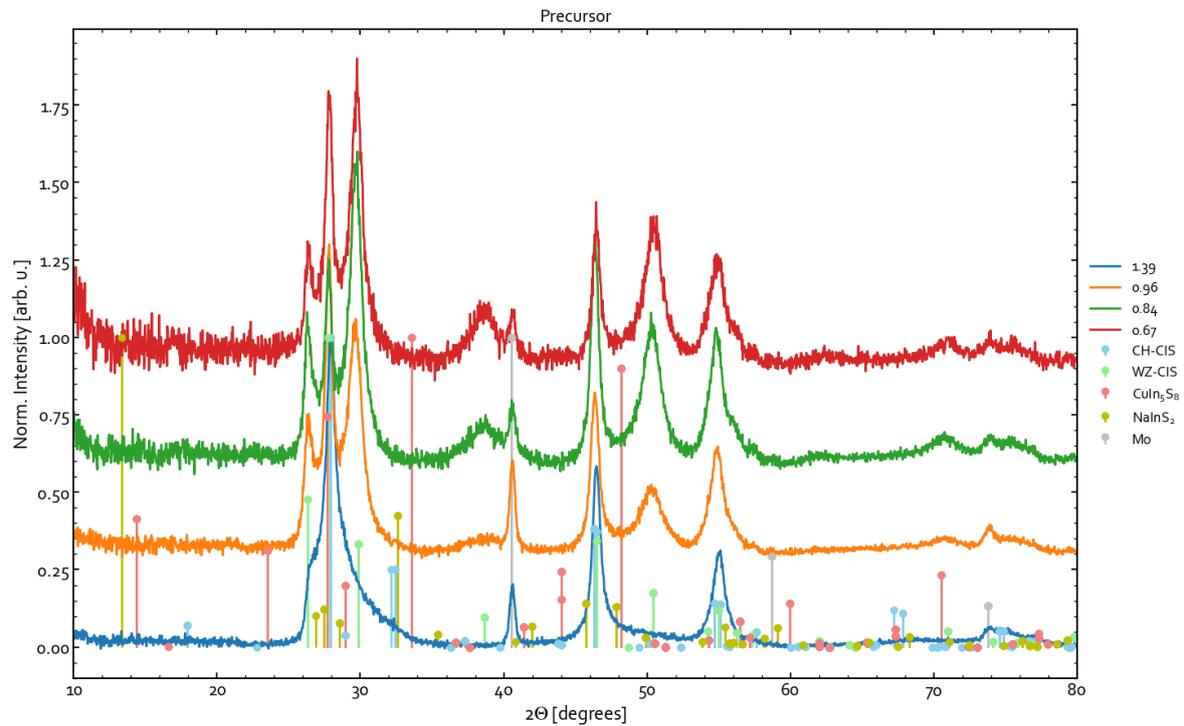

*Figure S 9: Complete GIXRD ($d_{inc}$ = 1°) of Cu-In-S precursors with various compositions as indicated by [Cu]/[In] in the legend. The reference patters originate from the ICSD as described in the main paper.*

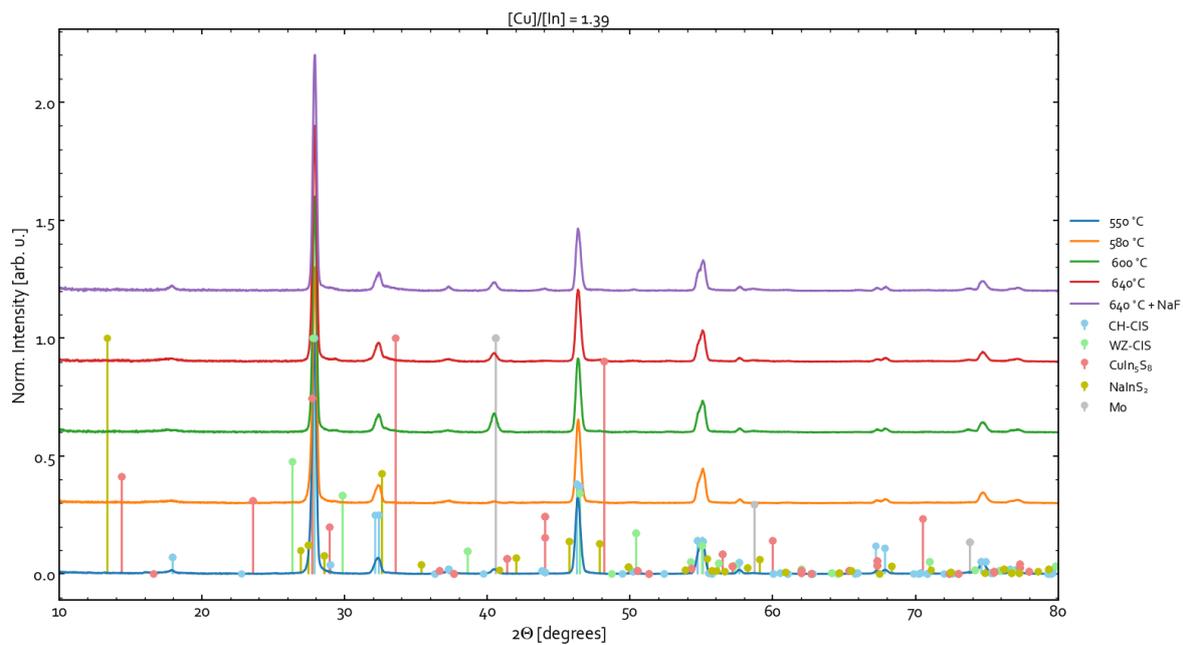

*Figure S 10: Complete GIXRD ($d_{inc}$ = 1°) of Cu-In-S samples with [Cu]/[In] = 1.39, annealed at various temperatures. The reference patterns originate from the ICSD as described in the main paper.*

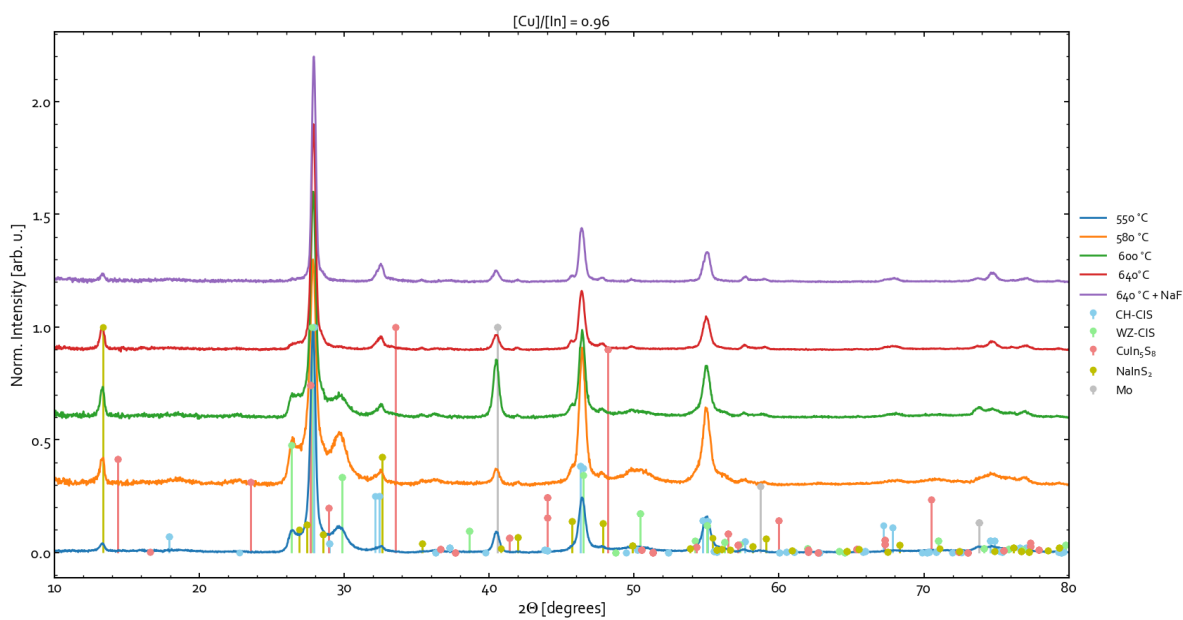

*Figure S 11: Complete GIXRD ($d_{inc}$ = 1°) of Cu-In-S samples with [Cu]/[In] = 0.96 annealed at various temperatures. The reference patterns originate from the ICSD as described in the main paper.*

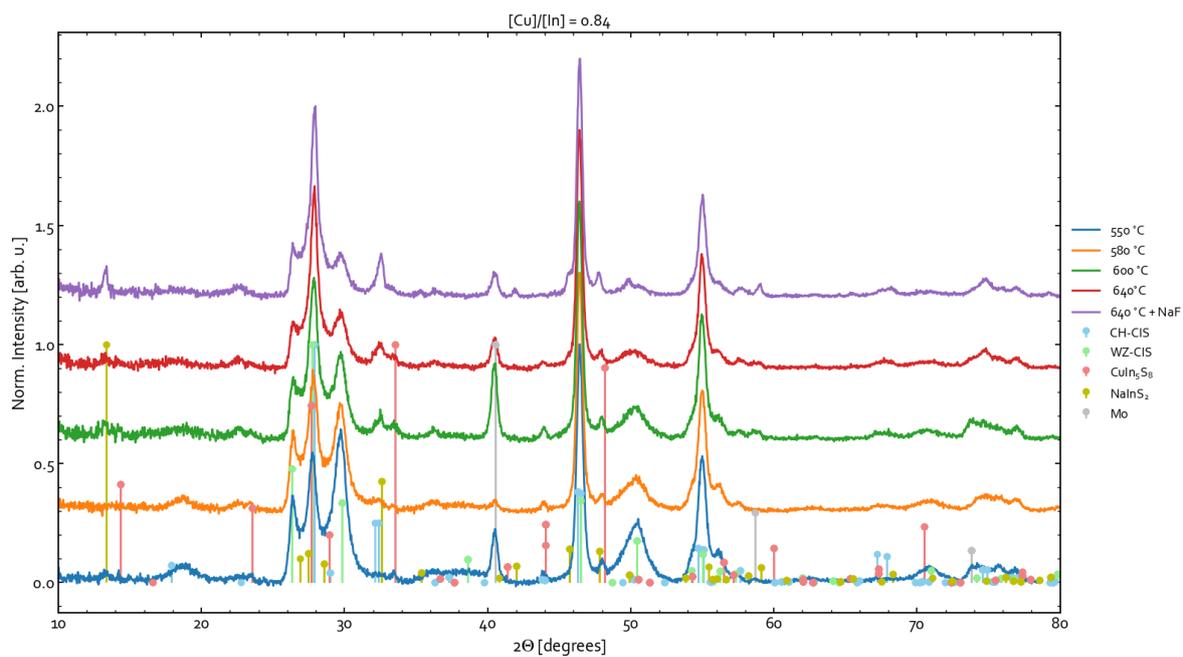

*Figure S 12: Complete GIXRD ($d_{inc}$ = 1°) of Cu-In-S samples with [Cu]/[In] = 0.84, annealed at various temperatures. The reference patterns originate from the ICSD as described in the main paper.*

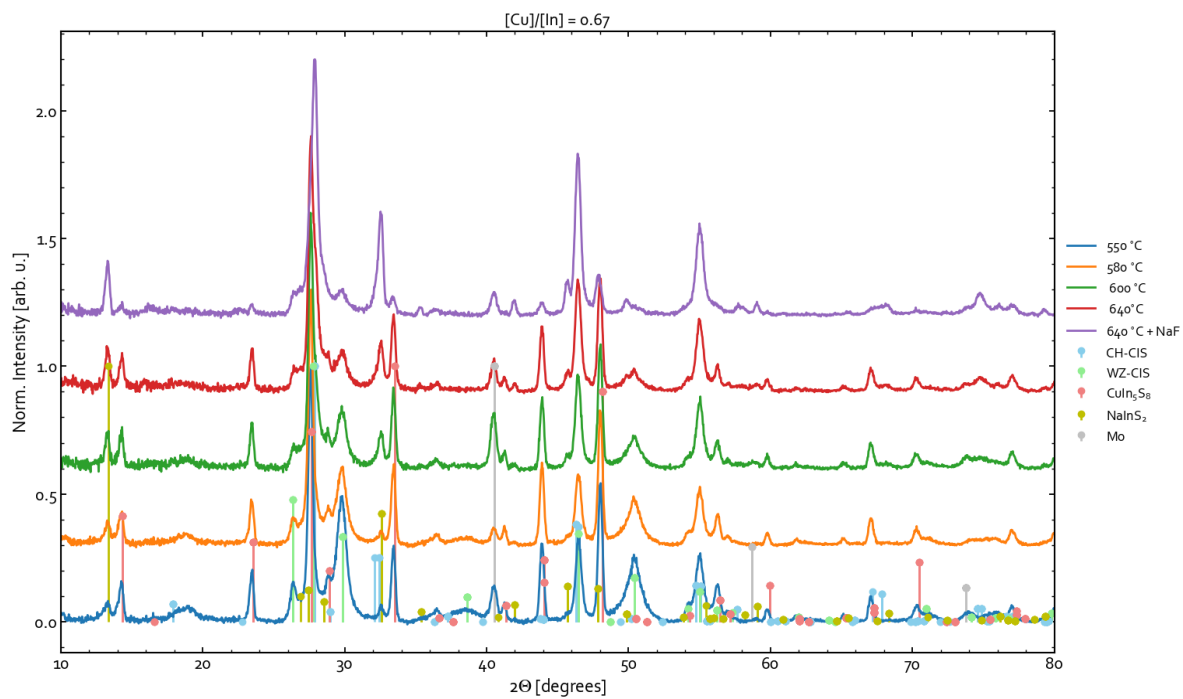

*Figure S 13: Complete GIXRD ($d_{inc}$ = 1°) of Cu-In-S sample with [Cu]/[In] = 0.67, annealed at various temperatures. The reference patterns originate from the ICSD as described in the main paper.*

# 3. Raman spectra of all samples
## 1.5. Precursors

The Raman spectra of the precursors measured with an excitation wavelength of 633 nm were presented in the main paper (Figure 3). For completeness, the in the spectra collected with excitation of 532 and 785 nm lasers are presented in Figure S 14 and Figure S 15.

### 1.5.1. Excitation wavelength of 532 nm

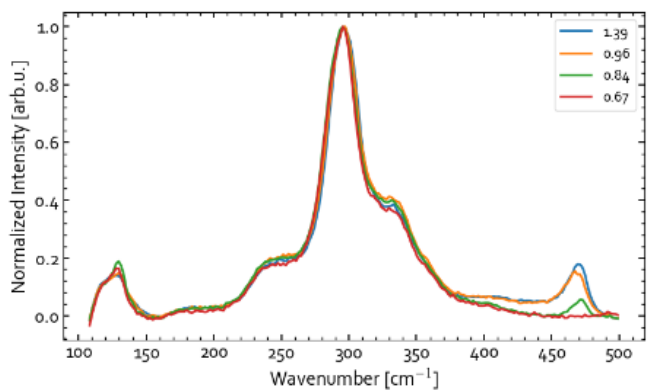

*Figure S 14: Raman spectra of Cu-In-S precursors with various compositions expressed as [Cu]/[In] ratios in the legend measured with $\lambda_{exc}$ = 532 nm.*

### 1.5.2. Excitation wavelength of 785 nm

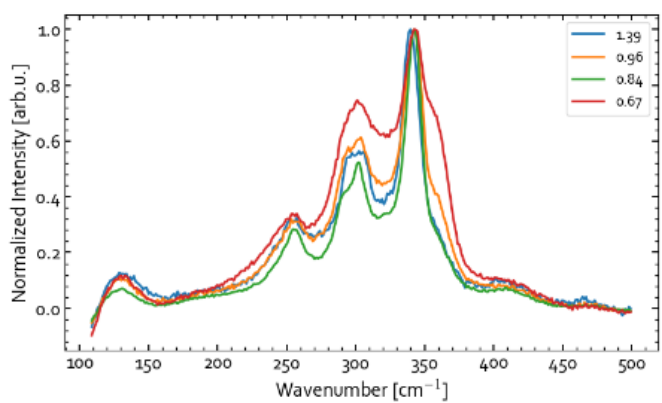

*Figure S 15: Raman spectra of Cu-In-S precursors with various compositions expressed as [Cu]/[In] ratios in the legend measured with $\lambda_{exc}$ = 785 nm.*

## 1.6. Raman of annealed samples

The Raman spectra of the samples in the main paper were measured with an excitation wavelength of 633 nm (Figure 6). For completeness and to illustrate the impact of the excitation wavelength on the Raman spectrum Figures S16-S23 presents the Raman spectra measured with 532 and 785 nm excitation.

### 1.6.1. Excitation wavelength of 532 nm

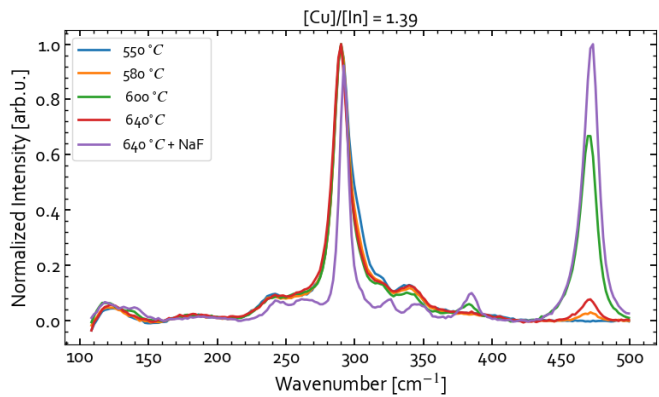

*Figure S 16: Raman spectra of CIS with [Cu]/[In] = 1.39 annealed at various temperatures measured with $\lambda_{exc}$ = 532 nm.*

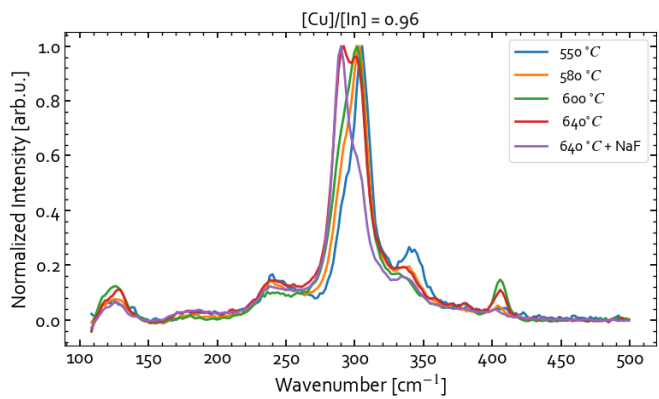

*Figure S 17: Raman spectra of CIS with [Cu]/[In] = 0.96 annealed at various temperatures measured with $\lambda_{exc}$ = 532 nm.*

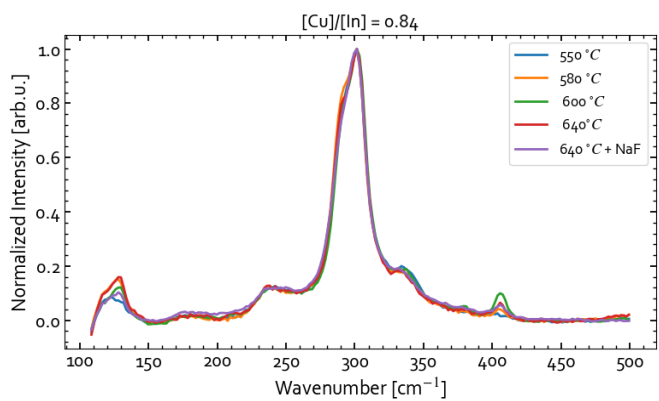

*Figure S 18: Raman spectra of CIS with [Cu]/[In] = 0.84 annealed at various temperatures measured with $\lambda_{exc}$ = 532 nm.*

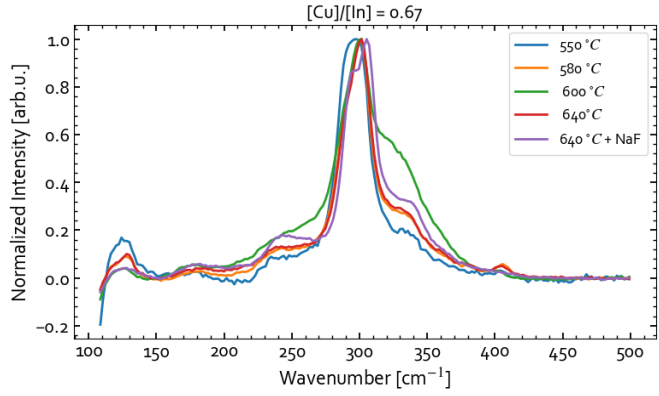

*Figure S 19: Raman spectra of CIS with [Cu]/[In] = 0.67 annealed at various temperatures measured with $\lambda_{exc}$ = 532 nm.*

### 1.6.2. Excitation wavelength of 785 nm

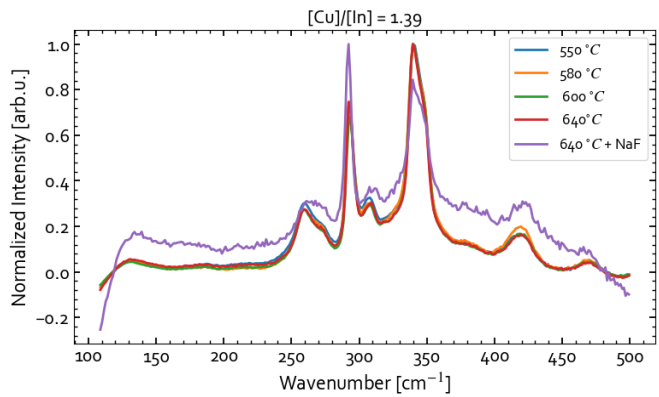

*Figure S 20: Raman spectra of CIS with [Cu]/[In] = 1.39 annealed at various temperatures measured with $\lambda_{exc}$ = 785 nm.*

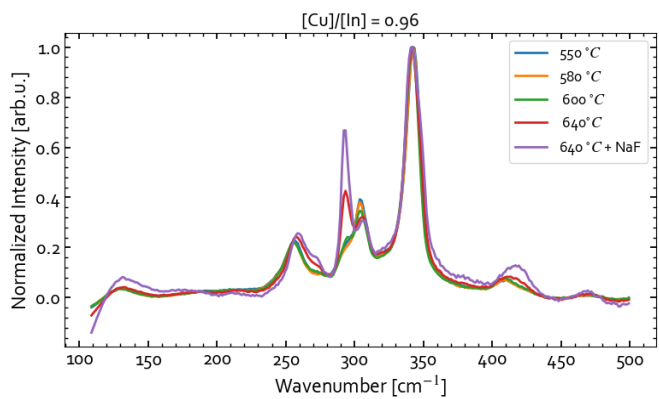

*Figure S 21: Raman spectra of CIS with [Cu]/[In] = 0.96 annealed at various temperatures measured with $\lambda_{exc}$ = 785 nm.*

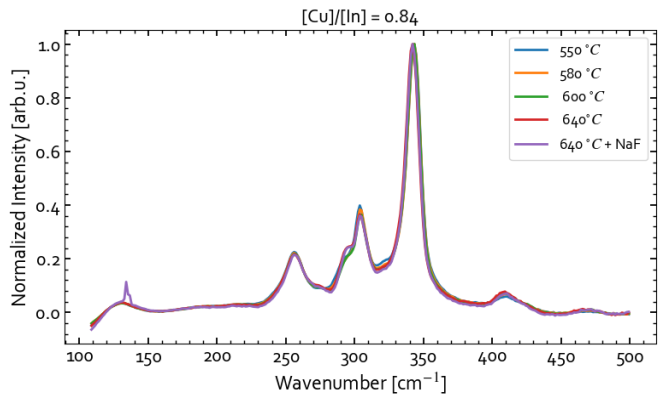

*Figure S 22: Raman spectra of CIS with [Cu]/[In] = 0.84 annealed at various temperatures measured with $\lambda_{exc}$ = 785 nm.*

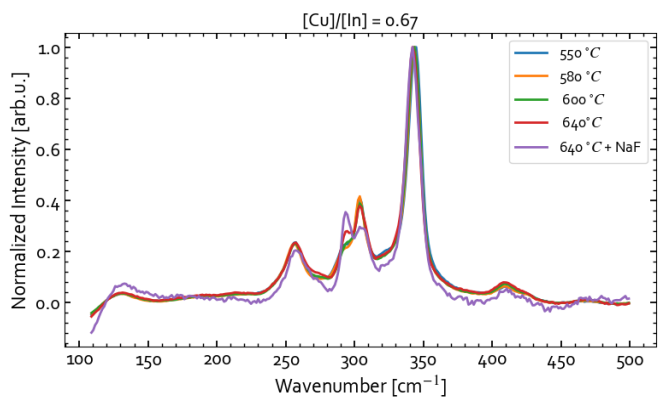

*Figure S 23: Raman spectra of CIS with [Cu]/[In] = 0.67 annealed at various temperatures measured with $\lambda_{exc}$ = 785 nm.*

## 4. Correlation of WZ-CIS phase measured by GIXRD and Raman spectroscopy

Figure S 24 shows the correlation between the peak ratios of XRD and the Raman spectra measured on the same samples. The ratio of XRD peaks ascribed to CH-CIS *(32.3°)* and WZ-CIS *(50.4°)* are plotted versus the ratio of Raman peak intensities at 340 cm$^{-1}$ and the *CH-CIS* A$_1$ mode at 292 cm$^{-1}$. The Raman spectra were measured with 633 nm excitation. There is a clear correlation between the WZ-phase measured by XRD and the 340 cm$^{-1}$ peak measured with Raman. (see Figures S20-24).

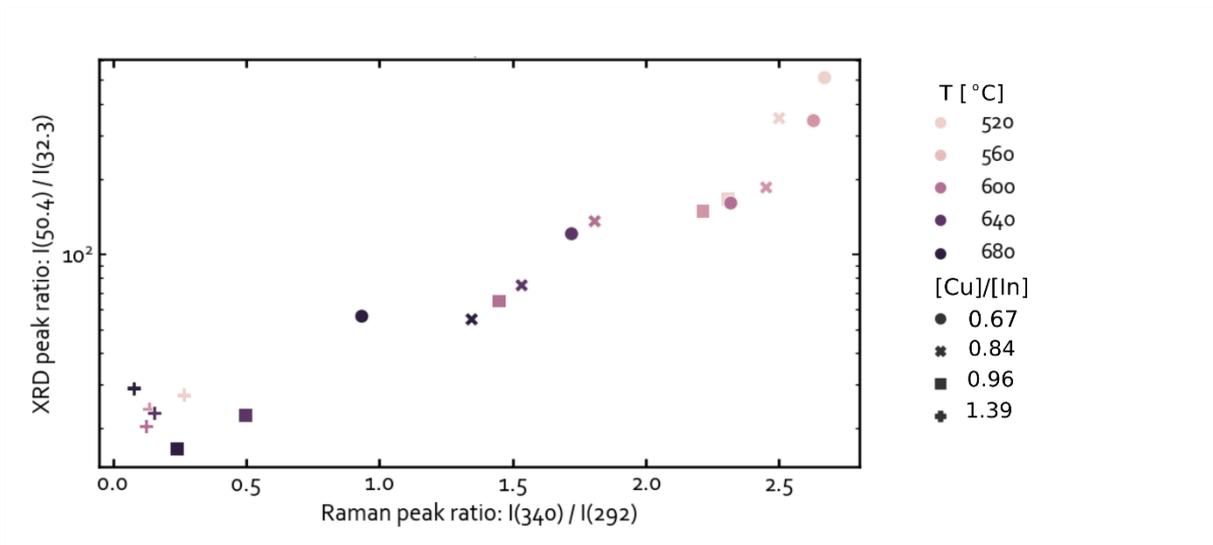

*Figure S 24: Correlation of XRD and Raman characteristics. The ratio of XRD peak intensities at I(50.4)/I(32.5) is a measure of the WZ-CIS phase content relative to CH-CIS phase content. The correlation with the Raman peak ratio I(340)/I(292), where I(292) is the $A_1$ mode of CH-CIS, indicates that the 340 cm$^{-1}$ peak relate to WZ-CIS.*

## 5. Formation enthalpy of Cu-poor Cu-In-S structures

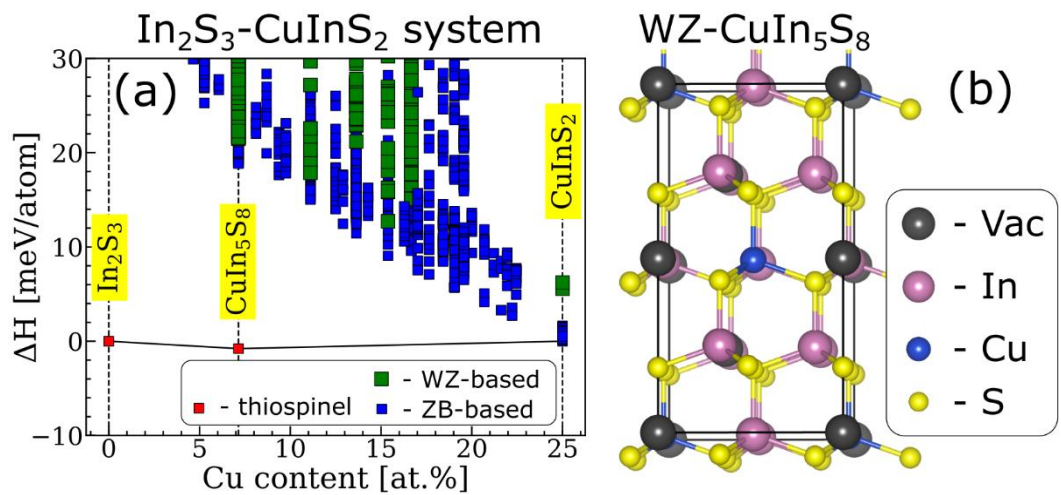

*Figure S 25: (a) Convex hull constructed for $In_2S_3$-$CuInS_2$ pseudo-binary system based on the thiospinel structures of $CuIn_5S_8$ and $In_2S_3$ from literature, as well as ZB- and WZ-based polytype structures generated using our screening algorithm. (b) The most stable WZ-based $CuIn_5S_8$ structure discovered. Details of the screening algorithm will be published elsewhere shortly.*